\documentclass[twocolumn]{aastex631}

\graphicspath{{./}{}}
\usepackage{gensymb}
\usepackage{graphicx}
\usepackage{amsmath}
\usepackage{booktabs}
\usepackage{multirow}

\usepackage{calligra}
\DeclareMathAlphabet{\mathcalligra}{T1}{calligra}{m}{n}
\DeclareFontShape{T1}{calligra}{m}{n}{<->s*[2.5]callig15}{}

\PassOptionsToPackage{dvipsnames,usenames}{xcolor}
\definecolor{BlueGreen}{HTML}{00B3B8}

\newcommand\AH{{AH76}}

\newcommand{\dva}[1]{{#1}}

\makeatletter
\DeclareRobustCommand{\volume}{\text{\volumedash}V}
\newcommand{\volumedash}{%
  \makebox[0pt][l]{%
    \ooalign{\hfil\hphantom{$\m@th V$}\hfil\cr\kern0.08em--\hfil\cr}%
  }%
}

\begin{document}
\title{On Binary Formation from Three Initially Unbound Bodies}

\author{Dany Atallah}
\affiliation{Department of Physics \& Astronomy, Northwestern University, Evanston IL 60208, USA}
\affiliation{Center for Interdisciplinary Exploration \& Research in Astrophysics (CIERA), Evanston, IL}

\author{Newlin C.\ Weatherford}
\affiliation{Department of Physics \& Astronomy, Northwestern University, Evanston IL 60208, USA}
\affiliation{Center for Interdisciplinary Exploration \& Research in Astrophysics (CIERA), Evanston, IL}

\author{Alessandro A.\ Trani}
\affiliation{Niels Bohr International Academy, Niels Bohr Institute, Blegdamsvej 17, 2100 Copenhagen, Denmark}
\affiliation{Research Center for the Early Universe, School of Science, The University of Tokyo, Tokyo 113-0033, Japan}
\affiliation{Okinawa Institute of Science and Technology, 1919-1 Tancha, Onna-son, Okinawa 904-0495, Japan}

\author{Frederic A.\ Rasio}
\affiliation{Department of Physics \& Astronomy, Northwestern University, Evanston IL 60208, USA}
\affiliation{Center for Interdisciplinary Exploration \& Research in Astrophysics (CIERA), Evanston, IL}

\begin{abstract}
We explore three-body binary formation (3BBF), the formation of a bound system via gravitational scattering of three initially unbound bodies (3UB), using direct numerical integrations. For the first time, we consider systems with unequal masses, as well as finite-size and post-Newtonian effects. Our analytically derived encounter rates and numerical scattering results reproduce the 3BBF rate predicted by Goodman \& Hut (1993) for hard binaries in dense star clusters. We find that 3BBF occurs overwhelmingly through nonresonant encounters and that the two most massive bodies are never the most likely to bind. Instead, 3BBF favors pairing the two least massive bodies (for wide binaries) or the most plus least massive bodies (for hard binaries). 3BBF overwhelmingly favors wide binary formation with super-thermal eccentricities, perhaps helping to explain the eccentric wide binaries observed by Gaia. Hard binary formation is far rarer, but with a thermal eccentricity distribution. The semimajor axis distribution scales cumulatively as $a^3$ for hard and slightly wider binaries. Though mergers are rare between black holes when including relativistic effects, direct collisions occur frequently between main-sequence stars---more often than hard 3BBF. Yet, these collisions do not significantly suppress hard 3BBF at the low velocity dispersions typical of open or globular clusters. Energy dissipation through gravitational radiation leads to a small probability of a bound, hierarchical triple system forming directly from 3UB. 
\end{abstract}


\section{Introduction} \label{sec:intro}

The formation of binaries containing stellar and compact objects is essential to the production of numerous high-energy astrophysical phenomena, including gravitational wave emission and/or fast radio bursts released with compact object mergers \citep[e.g.,][]{Rodriguez_2019, Kremer_2021}, X-ray binaries \citep[][]{Sana_2012}, and supernovae \citep[][]{Maoz_2014}. Binaries are also essential to the evolution of dense stellar environments since they act as dynamical heat sources that expand the cluster's core through repeated scattering interactions---\textit{binary burning} \citep[e.g.,][]{Heggie_2003}---and promote stellar collisions and tidal disruption events \citep{Bacon1996, Fregeau2004, Ryu_2023}.

Many stellar binaries form `primordially' in molecular clouds \citep[e.g.,][]{Shu_1987}, but also dynamically from two fully formed and isolated bodies, especially in dense stellar environs. Several types of dissipative effects may bind two lone stars together, including dynamical friction in a gaseous medium \citep{Rozner_2023}, tidal heating of one star by another \citep[`tidal capture'; e.g.,][]{Fabian_1975, Generozov_2018}, and gravitational wave emission in a close passage of two compact objects \citep[`gravitataional wave capture'; e.g.,][]{Quinlan_1989}. In this work, we explore a purely Newtonian phenomenon, three-body binary formation (3BBF), in which three isolated (energetically unbound) bodies pass near each other and gravitationally scatter to form a new binary. The leftover single acts as a source of dissipation in this scenario, a catalyst, transferring gravitational potential energy into the kinetic energies of the single and a new binary's center of mass.

The body of work investigating 3BBF is presently very limited compared to investigations of the aforementioned binary formation mechanisms. The historical lack of interest is likely due to over-generalization of early analytic estimates of 3BBF's impact on star clusters \citep{Heggie_1975,Stodolkiewicz1986,Goodman_1993}. The usual narrative states that the 3BBF rate is negligible over most of a cluster's dynamical lifetime, except in the short window of time central densities spike during the core collapse process \citep[e.g.,][]{Hut1985,FreitagBenz2001,Joshi2001} or even thereafter \citep[e.g.,][]{Statler1987,Hut1992}.\footnote{Here, \textit{core collapse} is the process by which the most massive objects in a stellar cluster rapidly evacuate kinetic energy from central regions, sink deeper into the cluster (dynamical friction), and subsequently contract the core to increasingly higher densities, a process that is halted by 3BBF.} Such studies generally predate the cluster modeling community's widespread incorporation of primordial binaries and realistic IMFs---and therefore neglect essential black hole (BH) dynamics. Because of the 3BBF rate's steep mass dependence, ${\propto}\, n^3 G^5 m^5 \sigma^{-9}$ \citep[e.g.,][]{Heggie_1975,Kulkarni1993,OLeary2006,Banerjee2010,Morscher_2013,Morscher_2015}, BH populations greatly enhance 3BBF. Without such massive bodies, efficient 3BBF would require an extreme cluster density only achieved in very deep core collapse even beyond the central densities of today's \textit{observationally} `core-collapsed' Milky Way globular clusters, whose small cores remain supported by central binary burning \citep[e.g.,][]{Kremer_2021_WDs}. In fact, accounting for BH retention and using the 3BBF recipe of \citet{Morscher_2015}, \citet{Weatherford_2023a} predict that 3BBF occurs frequently in globular clusters---an overwhelming majority involving BHs---and cumulatively powers many high-speed ejections. Given its impact on the formation of dynamically active BH binaries and high-speed ejections, 3BBF is essential for understanding star cluster evolution, BH mergers, and high-velocity stellar populations.

Earlier disinterest in 3BBF may also stem from its overwhelming tendency to form soft binaries, coupled with the assumption that such binaries are unlikely to survive long enough to contribute significantly to cluster dynamics via binary burning. Indeed, strong encounters quickly disrupt most soft binaries and also tighten those formed especially hard until they merge or are ejected from the cluster \citep[e.g.,][]{HutInagaki1985,McMillan1986,GoodmanHernquist1991,Bacon1996,ChernoffHuang1996,Fregeau2004}. This reasoning was previously used to justify neglecting 3BBF, either entirely \citep[e.g.,][]{Joshi2000,Fregeau2003}, or if none of the bodies were BHs \citep[e.g.,][]{Morscher_2015}, in prescription-based Monte Carlo star cluster models such as \texttt{CMC} \citep{CMCRelease}. Unfortunately, this neglects the formation of binaries of only moderate hardness, with or without BHs. A 
striking result of \citet{Goodman_1993} is that soft binaries from 3BBF, though typically short-lived, form so frequently that the small fraction which \textit{do} survive and harden sufficiently may yield over $90\%$ of hard binaries over long timescales in massive star clusters. These binaries would, in fact, survive long enough to contribute substantially to binary burning but are not typically accounted for in cluster modeling (outside of direct $N$-body simulations, which have their own limitations; see below).

Despite renewed interest, modern 3BBF recipes \citep[e.g.,][]{Goodman_1993,Ivanova_2005,Ivanova_2010,Morscher_2013,Morscher_2015} are untested by numerical scattering experiments. The only two examples of such studies, \citet{Agekyan_1971} and \citet[][]{Aarseth_1976}---hereafter referenced as \AH{}---suffered from small sample sizes and were limited to equal point masses. And while full direct $N$-body codes capture 3BBF naturally, a detailed analysis of this physical process is challenging due to the rarity of the event in the low-mass, small-$N$, or low-density clusters typically modeled by such codes \citep{Tanikawa_2013,Pina_2023}. 

To this day, the direct $N$-body approach borders on being too computationally expensive to practically simulate globular clusters that are simultaneously as massive, old, \textit{and} dense as those in the Milky Way \citep{Wang2016,ArcaSedda2023}. Conversely, much faster Monte Carlo and semi-analytic codes use highly approximate recipes \citep[e.g.,][and references therein]{CMCRelease} for 3BBF rather than direct integration with a small-$N$-body code---e.g., \textsc{Fewbody} \citep{Fregeau2004} or \textsc{Tsunami} \citep{Trani2023}. \texttt{CMC}'s prescription for 3BBF first parses the cluster's radially sorted list of bodies (excluding binaries) into sets of three. For each such set, it then decides in probabilistic Monte Carlo fashion whether or not to pair the two most-massive bodies---ignoring the possibility of pairing the lowest mass body---in the set based on the local 3UB encounter rate, estimated from the bodies' masses and the local mean stellar density and velocity dispersion \citep[see Section~2.3.1 of][]{CMCRelease}. Importantly, \texttt{CMC}'s 3BBF prescription has only cursorily been tested against direct $N$-body simulations \citep[][]{Morscher_2013}, in which 3BBF occurs naturally; the accuracy of \texttt{CMC}'s 3BBF rate is therefore uncertain. In particular, no rigorous justification exists for the choice to pair the two most-massive bodies in each 3UB encounter---a choice that may significantly affect newly formed binaries and cluster evolution.

Here, we present a rigorous framework for 3UB interactions and self-consistently investigate 3BBF physics through direct $N$-body scattering experiments. Our methodology builds upon the work done by \AH{} with adjustments made to correct a minor inconsistency in the Monte Carlo sampling scheme \AH{} adopted from \citet{Agekyan_1971}. Our investigation is built on the \textsc{Tsunami} integrator (\citealt{Trani2023}; Trani, Spera, \& Atallah 2024, in preparation) and the \textsc{CuspBuilding} Python package \citep{Atallah_2023}.

We describe our methodology in Section~\ref{sec:methods}, explaining the initial condition algorithm in Sections~\ref{sec:IC_Alg} and \ref{sec:sec:termination_of_IC}. Section~\ref{sec:encounter_rate_derivation} features a first-principles derivation of the 3UB encounter rate complementary to our algorithm, validated to be correct to within percent error in the particle-in-box simulation. We reproduce the results of \AH{} using their 3UB algorithm in Section~\ref{sec:reproducing_AH} and also justify the need for correcting the original \AH{} algorithm by analyzing a simple particle-in-box simulation. In Section~\ref{sec:pointmass3BBF}, we broadly explore 3BBF in the point mass limit. The equal-mass hard binary formation rate predicted by \citet{Goodman_1993} is reproduced in Section~\ref{sec:hard_binary_formation} and we discuss our findings regarding super-thermal wide binary formation in Section~\ref{sec:widebinaries}. We investigate the 3UB scattering of unequal point masses in Section~\ref{sec:unequal_mass} before applying our framework to the scattering of bodies with finite-size, main-sequence (MS) stars and black holes (BHs) in Section~\ref{sec:finitesize_PN}.  We lay out our conclusions and discuss next steps in Section~\ref{sec:conclusion}.

\section{Methods} \label{sec:methods}
\dva{We conduct our three-body scattering experiments with the \textsc{CuspBuilding} Python package \citep{Atallah_2023}, a Monte Carlo scattering framework built upon the \textsc{Tsunami} integrator (\citealt{Trani2023}; Trani, Spera, \& Atallah 2024, in preparation). \textsc{Tsunami} is a direct $N$-body integrator based on Mikkola's algorithmic regularization \citep{tsunami2022,trani2019a,trani2019b}, using the leapfrog algorithm in conjunction with Bulirsch-Stoer extrapolation \citep{stoer1980} and the chain-coordinate system introduced in \citet{kschain}. 

These techniques allow \textsc{Tsunami} to follow close encounters with extreme accuracy without reducing the integration timestep, unlike more traditional integrators used for stellar scattering calculations \citep[e.g.,][]{2004MNRAS.352....1F}. This makes \textsc{Tsunami} an ideal code for integrating any compact few-body system, including extreme mass-ratio configurations, such as stellar-mass binary BH scattering in the vicinity of an SMBH \citep{Trani_2023}. The \textsc{Tsunami/CuspBuilding} framework yields extreme precision and speed, with a typical evaluation rate of ${\sim}200$ 3UB scatterings per second, per CPU core. In total, we generate over $10^{10}$ 3UB encounters, one of the largest sets of scattering interactions yet generated for a single work.}

As in \citet{Agekyan_1971} and \AH{}, we initiate all bodies relative to the origin of an inertial reference frame, $O$. This origin serves as the ``target'' of all three bodies. Unlike in \AH{}, we adopt a ``spherical'' initial condition sampling method in contrast to their ``cylindrical'' method; we elaborate on this distinction and provide a robust numerical justification for adjusting this algorithm in Sections~\ref{sec:IC_Alg} and \ref{sec:reproducing_AH}.

\begin{figure}
\includegraphics[width=\columnwidth]{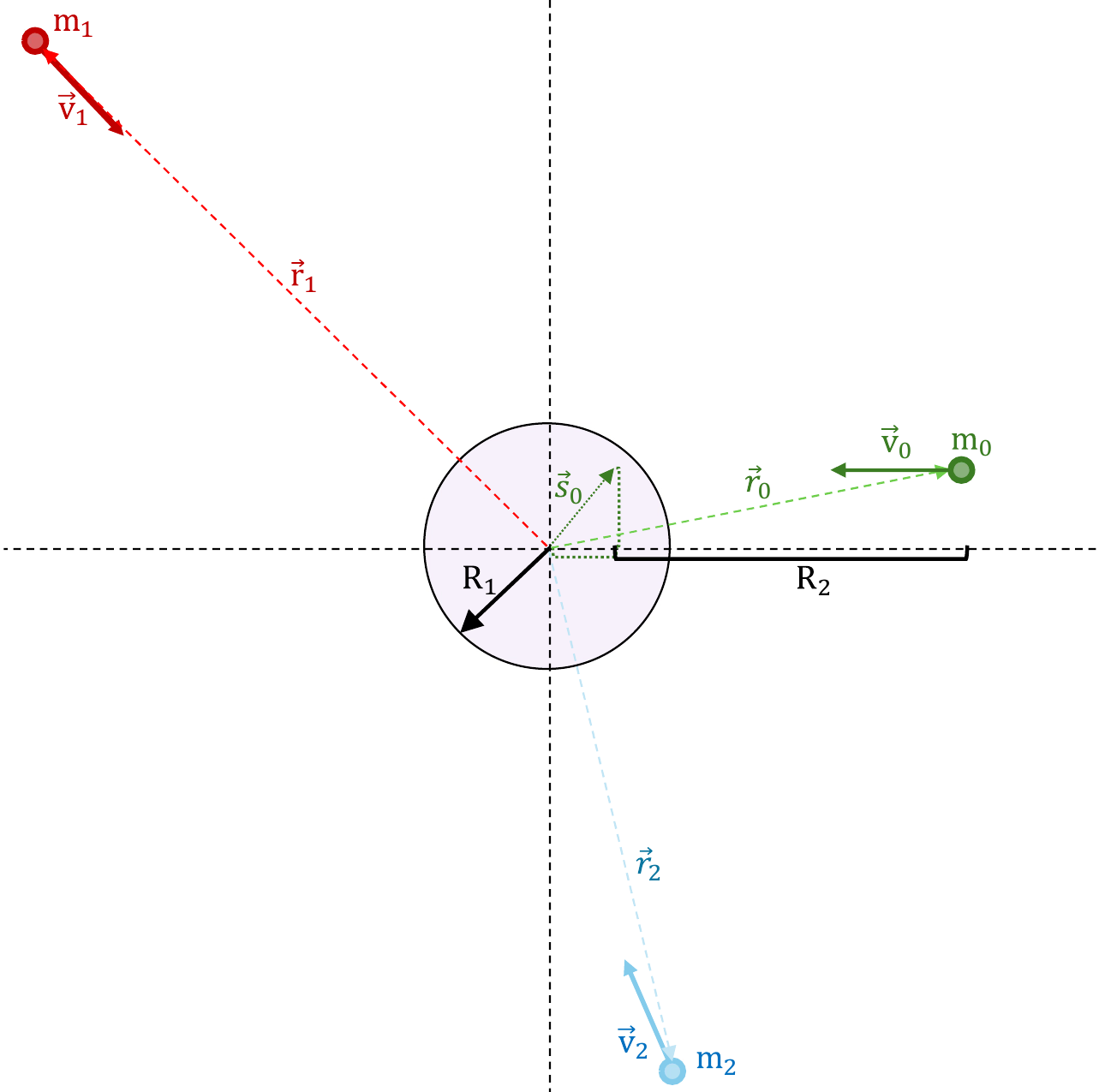}
\caption{\dva{A schematic of our three unbound body (3UB) initial condition algorithm, projected into two dimensions. The labels used here align with Equations~(\ref{eq:impact_parameter})--(\ref{eq:finalIC}).} }
\label{fig:ICschematic}
\end{figure}

\subsection{Initial Condition Algorithm}\label{sec:IC_Alg}
The initial condition algorithm may be subdivided into three parts:
\begin{enumerate}
    \item Select the masses and velocity vectors of the three scattering bodies.
    \item Randomly assign a point in a sphere of radius $R_1$ to each body. This point is drawn from a distribution explicitly uniform in the volume of a sphere.
    \item Pull each body backward in time along a straight line using their individual velocity vectors. This procedure approximates gravitationally isolating the three bodies far from the region of interaction.  
\end{enumerate}

We begin by defining a set of initial properties for each body: initial masses, $m_{\rm i}$, and velocity magnitudes, $v_{\rm i}$, with $\rm{i} = \{0,1,2\}$. For all scattering experiments in this work, we choose to randomly sample the velocities of all three bodies from a single  shared Maxwellian distribution. We make this choice to reduce the parameter space of initial conditions for our experiments, but note that sampling velocities from three separate distributions may be more realistic in cases where the bodies have unequal masses (e.g., Section~\ref{sec:unequal_mass}) and progression towards partial energy equipartition is assumed in the local environment.

Each body is then assigned a position vector $\vec{s}_{\rm i}$ relative to the origin $O$ and velocity vector $\vec{v}_{\rm i}$ (both randomly sampled and isotropic) using the following relations:
\begin{equation}\label{eq:impact_parameter}
    \begin{aligned}
        \vec{s}_{\rm{i}} =& \, s_{\rm i} \vec{e}^{\, r}_{\rm i}, \qquad\vec{v}_{\rm{i}} = v_{\rm i} \vec{e}^{\, v}_{\rm i}, \\
        s_{\rm{i}} =& \,\sqrt[3]{{\cal U_{\rm i}}(0,1)}\,R_1, \qquad
        R_1 = \,2 \chi_1 b_{90}, \\ b_{90} =& \max\left[\frac{G (m_{\rm i}+m_{\rm j})}{\langle (\vec{v}_{\rm ij})^{2}\rangle}\right], \quad \vec{v}_{\rm ij} = \,\vec{v}_{\rm i} - \vec{v}_{\rm j},\\
        \vec{e}^{\, r}_{\rm i} =& \begin{bmatrix}
           \cos(\theta_{\rm i}) \sin(\phi_{\rm i}) \\
           \sin(\theta_{\rm i}) \sin(\phi_{\rm i}) \\
           \cos(\phi_{\rm i})\end{bmatrix},\; \;
           \vec{e}^{\, v}_{\rm i} = \begin{bmatrix}
           \cos(\alpha_{\rm i}) \sin(\beta_{\rm i}) \\
           \sin(\alpha_{\rm i}) \sin(\beta_{\rm i}) \\
           \cos(\beta_{\rm i})
         \end{bmatrix}.
    \end{aligned}
\end{equation}
Here, ${\cal U}(0,1)$ indicates a random sample from the uniform distribution between $0$ and $1$, $R_1$ is the radius of a region we call the \textit{interaction volume} centered on $O$, and $\chi_1$ is a coefficient we shall later vary to control the characteristic strength of the three-body encounter; $\chi_1$ dictates the size of $R_1$ in terms of $b_{90}$, the largest impact parameter between any of the possible two-body combinations that would yield a $90\degree$ deflection in an isolated two-body encounter.\footnote{Note that $\langle (\vec{v}_{\rm ij})^2\rangle = 3 \left(\sigma_{\rm i}^2 + \sigma_{\rm j}^2 \right)$ in the case of two bodies with velocities $(\vec{v}_{\rm i},\vec{v}_{\rm j})$ sampled from separate Maxwellians with one-dimensional velocity dispersions ($\sigma_{\rm i},\sigma_{\rm j}$), respectively \citep[e.g.,][]{Binney_2008}. Our assumption that the motions of all bodies are described by the same Maxwellian distribution thus corresponds to setting $\sigma_{\rm i } = \sigma_{\rm j }$, making $\langle (\vec{v}_{\rm ij})^2\rangle = 6 \sigma^2$.} In that spirit, we will often refer to $\chi_1$ as a dimensionless impact parameter as it serves a similar function to the impact parameter in traditional binary--single scattering experiments. 

To isotropically distribute $\vec{s}_{\rm{i}}$ and $\vec{v}_{\rm{i}}$ (and their corresponding unit vectors $\vec{e}^{\, r}_{\rm i}$ and $\vec{e}^{\, v}_{\rm i}$), we individually sample angles $(\theta_{\rm i},\beta_{\rm i})$ from the distribution ${\cal U}(0,2\pi)$, and the angles $(\phi_{\rm i},\alpha_{\rm i})$ from the distribution $\cos^{-1} \left[{\cal U}\left(-1,1\right)\right]$, repeating this process for each body. Our procedure samples points within a sphere of radius $R_1$ from a distribution uniform in volume. In contrast, \AH{} explore the parameter space using the volumes of arbitrarily rotated, overlapping cylinders---with ${s_{\rm i}^2 = b_{\rm{i}}^2 + z_{\rm i}^2}$, impact parameters $b_{\rm{i}} = \sqrt{{\cal U}(0,1)}\,R_1$, and offsets ${z_{\rm i} = {\cal U}(-1,1)\, R_1}$.

The set of points sampled by either procedure are \textit{not} the initial conditions for the scattering experiment. Rather, they describe positions at a point in time when all three bodies, if they were to travel at constant velocity on straight-line trajectories, would be within the interaction volume simultaneously. Like \AH{}, we refer to this point in time as the \textit{epoch time}, $t_{\rm e}$. If we were to initialize all bodies within the interaction volume, it would be likely (depending on $\chi_1$) that at least two bodies would already be energetically bound. Yet, we are interested in the formation of \textit{new} binaries from interactions involving three \textit{initially unbound} bodies. While we could simply resample any initial conditions containing bound pairs---and in fact we do, as  described shortly---this may need to be done many times if the initial positions are too close to each other. This would severely truncate the initial energy distribution, artificially enhancing both the separations between bodies and their velocities at the epoch time. So, like \AH{}, we initialize bodies far from the interaction volume by pulling each body backwards in time along the straight line parallel to its velocity vector.

To pull back all bodies by at least a chosen distance $R_2$, we may define the epoch time in terms of the slowest body's velocity:
\begin{equation}
t_{\rm e} = \frac{R_2}{\min (v_{\rm i})},
\end{equation}
where we choose $R_2=15 R_1$ to be consistent with \AH{}. Each body's offset distance, drawn backward along the aforementioned straight-line trajectory, is then
\begin{equation}
    \Delta \vec{r}_{\rm i} = -\vec{v}_{\rm i} t_{\rm e} = -v_{\rm i} t_{\rm e} \vec{e}^{\, v}_{\rm i} .
\end{equation}
With these offsets in hand, the initial conditions for each body are\footnote{In \AH{}, the unit vectors are $\vec{s}_{\rm i} = b_{\rm i} \vec{e}^{\, \perp}_{\rm i} - z_{\rm i} \vec{e}^{\, \parallel}_{\rm i}$, with $\vec{e}^{\, \perp}_{\rm i}, \vec{e}^{\, \parallel}_{\rm i}$ representing unit vectors perpendicular and parallel, respectively, to the randomly sampled velocity vectors of each body. Using our notation, $\vec{e}^{\, \parallel}_{\rm i}\equiv\vec{e}^{\, v}_{\rm i}$.}
\begin{equation}\label{eq:finalIC}
\begin{aligned}
    \vec{r}_{\rm i} &= \vec{s}_{\rm i} + \Delta\vec{r}_{\rm i} \\
    &= s_{\rm i} \vec{e}^{\, r}_{\rm i} - v_{\rm i} t_{\rm e} \vec{e}^{\, v}_{\rm i},\\
    \vec{v}_{\rm i } &= v_{\rm i} \vec{e}^{\, v}_{\rm i}.
\end{aligned}
\end{equation}
An example schematic of a 3UB initial condition is displayed in Figure~\ref{fig:ICschematic}.

Although starting the bodies outside the interaction volume does not entirely negate the possibility that at least two of the bodies are initially bound to each other, this is true of only $0.7\%$ of initial conditions sampled with our choice of $R_2=15 R_1$ (and $\chi_1 = 10$; see Section~\ref{sec:choice_of_chi1}). Like \AH, we exclude such instances from our results to avoid contaminating our sample of 3UB interactions. Specifically, we throw out any scattering experiment that starts with any two-body pairing or the entire three-body system having a negative total energy in the three-body center-of-mass frame. 

We do not filter out 3BBF events in which one or more bodies never enter the interaction volume, contributing ${\approx}20\%$ of 3BBF events for our choice of $R_2=15 R_1$. That 3BBF events can occur without all bodies reaching the interaction volume is a direct consequence of gravitational scattering. If the bodies were to travel on constant-speed, straight-line trajectories, then the above initial conditions would \textit{guarantee} that all three bodies are in the interaction volume at time $t_{\rm e}$. In reality, however, bodies initialized sufficiently close to each other (even when not bound) may still interact strongly enough that one or more bodies miss the interaction volume. One would hope to account for this effect when selecting initial conditions, but the exact trajectories in the three-body problem cannot be determined analytically. 

The lack of analytic gravitational focusing prescriptions in 3UB interactions contrasts with traditional binary--single scatterings, which effectively become two-body hyperbolic encounters at sufficiently large separations---allowing scattering codes to easily incorporate two-body gravitational focusing when sampling initial conditions. A similar hyperbolic limit does not exist for 3UB interactions as each two-body separation increases proportionately to any increase in the starting distances relative to $O$. Thus, analogous gravitational focusing effects in 3UB interactions cannot be generically decomposed into separate two-body focusing terms without potentially biasing the experiment's final outcome. This renders futile efforts to generically pick initial conditions outside $R_1$ which simultaneously \textit{guarantee} that all three bodies must at some point cross within $R_1$. While unfortunate, drawing bodies backwards is necessary to mitigate potential biases in the initial energy distribution (caused by attempts to artificially modify initial conditions to predict 3UB gravitational focusing) or throwing away initial conditions that contain bound pairs.\footnote{Note that we have rigorously tested methods of setting up 3UB interactions other than the one employed in this work, such as targeting an incoming single at the center of mass of a hyperbolic two-body encounter---the setup often used in analytic estimates of the 3BBF rate. We find that, even for the case of equal masses and velocities, setups of this kind bias the pairing probability toward the incoming single. Our setup has no such bias.}

\subsection{Termination of 3UB Integration}\label{sec:termination_of_IC}
The duration of every scattering experiment (3UB interaction) is entirely adaptive and individualized to the experiment's specific initial conditions. \textsc{CuspBuilding} integrates the system until $t=2 t_{\rm e}$ before assessing whether all outgoing hierarchies---including instances of a single star unbound to any other star---are energetically isolated. Two hierarchies are energetically isolated from each other if their gravitational potential energy is less than $1\%$ of the kinetic energy of their relative motion (i.e., not including any internal binding energy of either hierarchy). If all hierarchies are energetically isolated from each other, then \textsc{CuspBuilding} considers the system to have reached its final state and terminates integration. 

Arithmetically stated, the specific energy of two outgoing hierarchies is $\epsilon_{\rm ij} = \frac{v_{\rm ij}^2}{2} - \frac{G \left(m_{\rm i} + m_{\rm j}\right)}{r_{\rm ij}}$, where $m_{\rm i}$ and $m_{\rm j}$ are the total masses of the two hierarchies, $\vec{v}_{\rm ij} = \vec{v}_{\rm i} - \vec{v}_{\rm j}$ is the relative velocity between their centers of mass, and $r_{\rm ij} = \left\| \vec{r}_{\rm i} - \vec{r}_{\rm j}\right\|$ is the distance between their centers of mass. We simply demand that $2 \epsilon_{\rm ij}/v_{\rm ij}^2>0.99$. Rewritten, each scattering calculation ends when $\frac{2 G \left(m_{\rm i} + m_{\rm j}\right)}{v_{\rm ij}^2 r_{\rm ij}}<0.01$ for all two-hierarchy combinations.  

\subsection{Encounter Rate}\label{sec:encounter_rate_derivation}
To aid comparison of our results to earlier literature on 3BBF, we here present the first \textit{encounter rate} that correctly predicts the probability per unit time of a 3UB interaction occurring. Note that a 3UB encounter rate is not equivalent to the 3BBF rate---i.e., the rate of binaries successfully forming from 3UB interactions. The latter is obtainable by multiplying the 3UB encounter rate by a numerically determined likelihood of 3BBF from a 3UB encounter, but earlier work \citep{Heggie_1975, Goodman_1993} instead estimated the 3BBF rate using detailed balance, bypassing a need for a 3UB encounter rate.

We consider two different geometric interpretations for the encounter rate:
\begin{itemize}
    \item A stationary spherical volume embedded in a host environment containing particle fields with local number densities, $n_{\rm i}$, and velocity dispersions, $\sigma_{\rm i}$, for each of up to three distinct populations $i\in[1,3]$.
    \item A spherical volume containing one target body---not necessarily at its center---that moves with the velocity of the target and thus has the same velocity dispersion, $\sigma_{\rm t}$, relative to its host environment.
\end{itemize}

To derive the encounter rate, we first define the probability of finding a body within an enclosing volume embedded in a particle field,
\begin{equation}
    P_{\rm i} = n_{\rm i} \volume_{\rm i},
\end{equation}
where $\volume_{\rm i}$ is an enclosing volume and $n_{\rm i} \volume_{\rm i} \ll 1$. The encounter rate is thus
\begin{equation}
    \Gamma_{\rm i} = \frac{dP_{\rm i}}{dt} = n_{\rm i} \frac{d\volume_{\rm i}}{dt},
\end{equation} 
where $d\volume_{\rm i}/dt$ is the volumetric flow-rate. Fundamentally, the one-dimensional flow-rate through a volume may be expressed as
\begin{equation}
\begin{aligned}
    \frac{d\volume_{\rm i}}{dt} &= \frac{d\volume_{\rm i}}{dz_{\rm i}} \frac{dz_{\rm i}}{dt} = A_{\rm i} v_{\rm i}
\end{aligned}
\end{equation}
where $A_{\rm i}$ is the projected cross-sectional area in the flow direction of a particle field, and $v_{\rm i}$ is the velocity of that fluid. If $P_{\rm i}$ is labeled as an \textit{event}, then the probability of three independent and simultaneous events, a 3UB event, is
\begin{equation}
\begin{aligned}
    P_{\rm 3B} &= \frac{1}{l!} P_{\rm i} P_{\rm j} P_{\rm k}\\
            &= \frac{n_{\rm i} n_{\rm j} n_{\rm k}}{l!}\volume_{\rm i} \volume_{\rm j} \volume_{\rm k},
\end{aligned}
\end{equation}
where $l!$ is the standard correction for joint Poisson distributions when particles are drawn from the same field (e.g., $l=2$ if two particles are selected from the same field).
The respective mean encounter rate is
\begin{equation}\label{eq:gamma3bpreav}
    \Gamma_{\rm 3B} \equiv \frac{n_{\rm i} n_{\rm j} n_{\rm k}}{l!}
    \left(v_{\rm i} A_{\rm i} \volume_{\rm j} \volume_{\rm k} + v_{\rm j} A_{\rm j} \volume_{\rm i} \volume_{\rm k} + v_{\rm k} A_{\rm k} \volume_{\rm i} \volume_{\rm j} \right).
\end{equation}
For all bodies to meet within the same sphere of radius $R_1$, we may set $\volume_{\{\rm i,j,k\}} = 4 \pi R_1^3/3 $ and $A_{\{\rm i,j,k\}} = \pi R_1^2 $. The mean encounter rate then becomes
\begin{equation}\label{eq:gamma_ijk}
    \Gamma_{\rm 3B} = \frac{3 \beta^3 }{4 \,l!}  R_1^8 n_{\rm i} n_{\rm j} n_{\rm k} (v_{\rm i} + v_{\rm j} + v_{\rm k}),
\end{equation}
where $\beta = \frac{4 \pi}{3}$, and the rate per unit volume is
\begin{equation} \begin{aligned} \label{eq:gammavol_ijk}
    \tilde{\Gamma}_{\rm 3B} \equiv \frac{\Gamma_{\rm 3B}}{\volume} &= \frac{3 \beta^2 }{4 \,l!}  R_1^5 n_{\rm i} n_{\rm j} n_{\rm k} \left(v_{\rm i} + v_{\rm j} + v_{\rm k}\right).
\end{aligned} \end{equation}

In the case of an isotropic Maxwellian velocity distribution for each particle field, with dispersion $\sigma_{\rm i}$, then
\begin{equation}\label{eq:gammamax_ijk}
        \tilde{\Gamma}_{\rm 3B} = \frac{3 \beta^2}{\sqrt{2 \pi} \, l!} R_1^5  n_{\rm i} n_{\rm j} n_{\rm k} \left(\sigma_{\rm i} + \sigma_{\rm j} + \sigma_{\rm k}\right).
\end{equation}
This rate assumes a small enough volume that the number density of the local field is roughly constant.

To find an encounter rate relative to an individual target already embedded within a spherical volume of radius $R_1$ (i.e., the \textit{per-body} rate), we may set ${P_{\rm k} = 1}$. Here, ${P_{\rm 3B}' = \frac{1}{l!} P_{\rm i} P_{\rm j}}$. Following the above calculation, the three-body encounter rate per body is then
\begin{equation}
    \Gamma_{\rm 3B}' = \frac{3 \beta^2}{4 \, l!} R_1^5 n_{\rm i} n_{\rm j} \left(v_{\rm i}' + v_{\rm j}'\right),
\end{equation}
where $v_{\rm i}'$ is the velocity of body $i$ relative to the target, body $k$. For Maxwellian velocity distributions,
\begin{equation}\label{eq:gammamax_ij}
\begin{aligned}
    \Gamma_{\rm 3B}' &=  \frac{3\beta^2}{\sqrt{2 \pi} \, l!} R_1^5 n_{\rm i} n_{\rm j} \left(\sigma_{\rm i}' + \sigma_{\rm j}'\right)\\
    \sigma_{\rm i}' &= \sqrt{\sigma^2_{\rm t} + \sigma^2_{\rm i}},
\end{aligned}
\end{equation}
where $\sigma_{\rm t}$ is the velocity dispersion of the Maxwellian from which the target's velocity is drawn.

\begin{figure}
\includegraphics[width=\columnwidth]{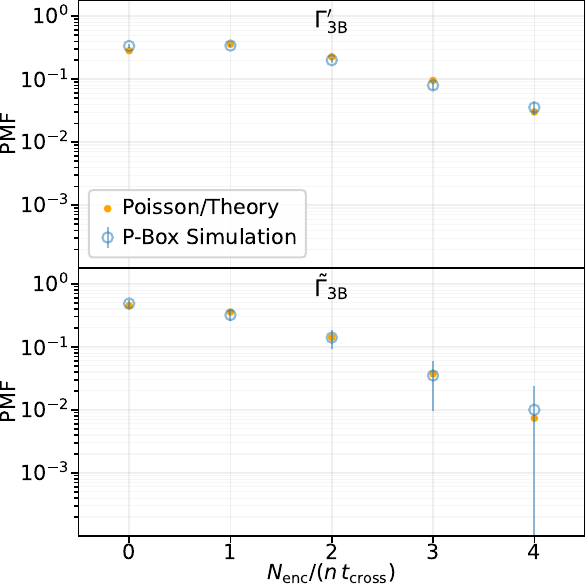}
\caption{
Yellow points show the discrete probability density function (PMF) for the number of times ($N_{\rm enc}$) three particles in a particle-in-box simulation are simultaneously within a small spherical volume at the box's center. The count on the horizontal axis is expressed as a rate by dividing $N_{\rm enc}$ by a large multiple, $n$, of the box's average crossing time, $t_{\rm cross}$, given particle speeds sampled from a Maxwellian distribution. The upper and lower panels show the per-body and volumetric encounter rates, with $n=5$ and $n=50$, respectively. Each particle travels at constant speed and elastically bounces off the walls of the box. The count $N_{\rm enc}$ increases by one each time a particle passes into the encounter volume while at least two other bodies are already inside the volume (or one other body in the per-body rate). The simulation results are compared to our derived encounter rates from Equation~(\ref{eq:gammamax_ijk})---blue points---by sampling from a Poisson distribution, ${P(N) = (\tilde{\Gamma}_{\rm 3B} \,\delta t)^N e^{-\tilde{\Gamma}_{\rm 3B} \, \delta t}/N!}$, where $\delta t = n \, t_{\rm cross}$.}
\label{fig:Encrate_PMF}
\end{figure}

To verify both the volumetric and per-body 3UB encounter rates, we conduct a simple particle-in-box simulation, with periodic boundary conditions, of non-interacting bodies on constant-velocity, straight-line trajectories (Figure~\ref{fig:Encrate_PMF}; see caption for further details). The 3UB encounter rates in the simulation are consistent with Equations~(\ref{eq:gamma_ijk})--(\ref{eq:gammamax_ij}) to within percent-level.

If all three particle fields have the same particle mass and Maxwellian velocity dispersion, then substituting Equation~(\ref{eq:impact_parameter}) into Equation~(\ref{eq:gammavol_ijk}) results in the scaling
\begin{equation}
     \tilde{\Gamma}_{\rm 3B} \propto \chi_1^5 \frac{G^5 m^5 n^3}{\sigma^9}.
\end{equation}
This reproduces the classic ${\sim}G^5 m^5 n^3\sigma^{-9}$ scaling from earlier estimates of the 3UB encounter rate \citep[e.g.,][]{Goodman_1993,Heggie_2003,Binney_2008}. However, not all 3UB encounters form binaries. So the true volumetric 3BBF rate, $\tilde{\Gamma}_{\rm F}$, must be numerically determined and satisfies the relation
\begin{equation}\label{eq:3BBF_vol_rate}
\tilde{\Gamma}_{\rm F} = P_{\rm ij} \tilde{\Gamma}_{\rm 3B},
\end{equation}
where $P_{\rm ij}$ is the numerically determined probability of forming a binary (with pairing $\{i,j\}$) per 3UB encounter occurring at rate $\tilde{\Gamma}_{\rm 3B}$.

\subsection{Replicating \& Assessing Aarseth \& Heggie 1976}\label{sec:reproducing_AH}
\begin{figure}
\includegraphics[width=\columnwidth]{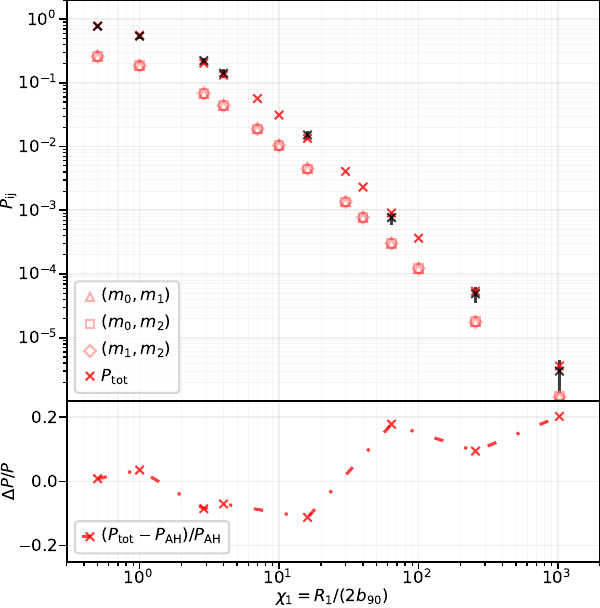}
\caption{Reproduction of three-body binary formation (3BBF) probabilities reported in \cite{Aarseth_1976}. \textit{Top Panel}: Probability $P_{\rm ij}$ of forming a binary containing bodies $i$ and $j$ as a function of dimensionless impact parameter $\chi_1$ (see Equation~\ref{eq:impact_parameter}) through the interaction of three equal-mass bodies initially on unbound trajectories. The initial velocities are drawn from a Maxwellian distribution. Red crosses mark the total 3BBF probability, independent of pairing, while shapes represent the pair-specific probabilities. As expected, equal-mass bodies have equal probabilities of pairing. Black crosses, with error bars, are the values reported in \AH{}. \textit{Bottom Panel}: Fractional difference in probabilities between our results and \AH{}. We see  good agreement $(\Delta{\lesssim}20\%)$.}
\label{fig:Pijvschi1_AH}
\end{figure}

As a final step before exploring new results, it is useful to reproduce the original 3BBF investigation conducted by \cite{Aarseth_1976} using their unmodified algorithm (i.e., with a cylindrically sampled impact parameter; see Section~\ref{sec:IC_Alg}) in the Newtonian, point mass regime. Figure~\ref{fig:Pijvschi1_AH} compares the 3BBF probability---the fraction of 3UB scattering experiments resulting in a binary forming---as a function of $\chi_1$ between AH76 and our \textsc{Tsunami} re-implementation of their method. As seen in the lower panel, our re-implementation results in no more than $\pm 20\%$ difference in the 3BBF probability from AH76, easily explained by the dramatic increase in sample size and computing resources. We also recover the $\chi_1^{-2}$ dependence of the total 3BBF probability and validate that the AH76 algorithm correctly results in indistinguishable bodies (with identical mass and velocities drawn from a single shared Maxwellian distribution) having an equal likelihood of pairing; see the red shapes in the top panel. 

The radial distribution of three bodies in a field occupying a fixed spherical volume cumulatively scales as ${C\left(r_{\rm i }\right) = \left(r_{\rm i}/R_1\right)^3}$, where $r_{\rm i}$ is the radial location of a body passing through a fixed volume of radius $R_1$. Unfortunately, this is not true of the initial condition algorithm employed by \AH{} which probes an asymmetric encounter volume comprised of randomly oriented overlapping cylinders. This expands the effective encounter volume by 50\%--150\% when compared to our desired spherical sampling algorithm and features irregular and non-linear behaviour for values of $r_{\rm i} \approx R_1$. Fewer binaries form as a result (see Section~\ref{sec:equalpointmasses}) and those that do are softer. 

\section{3BBF in the Point Mass Limit}\label{sec:pointmass3BBF}

We now explore the outcomes of 3UB scatterings for point-particles of both equal and unequal masses. To be consistent with our validated encounter rate, derived by invoking spherical symmetry, we use our spherical initial condition sampling algorithm  (Section~\ref{sec:IC_Alg}) instead of using the cylindrical prescription of \AH{}.

To aid astrophysical interpretation of our results, we first define our criteria for \textit{hard} and \textit{soft} binaries. Traditionally, this refers to binaries with binding energies greater than (hard) or less than (soft) the typical kinetic energy of gravitating bodies in the local environment \citep{Heggie_2003, Binney_2008}. We make a slight modification by using the \textit{fast} and \textit{slow} criteria justified by \citet{Hills_1990}, where the hardness of a newly formed binary is determined instead by its orbital velocity relative to the local velocity distribution surrounding the 3UB encounter. 
The \textit{fast/slow boundary} is
\begin{equation}\label{eq:fast_slow_boundary}
    a_{\rm fs,ij} = \frac{G (m_{\rm i} + m_{\rm j})}{\langle v_{\rm rel}^2\rangle},
\end{equation}
where ${m_{\rm i}, m_{\rm j}}$ are the masses of the two newly paired bodies and $\langle v_{\rm rel}^2\rangle$ is the typical mean-squared relative velocity of the new binary, simplified to be ${\langle v_{\rm rel}^2\rangle = 3 \sigma_{\rm rel}^2 = 6 \sigma^2}$ for an isotropic gas described by a Maxwellian with a one-dimensional velocity dispersion $\sigma$. Note that ${\max \left(a_{\rm fs,ij} \right)=b_{90}}$.

\dva{The definition of $a_{\rm fs,ij}$ and $b_{90}$ are dependent on the velocity dispersion selected for a set of scattering experiments. In effect, we may extract scale-free results in the Newtonian regime simply by dividing 3UB length scales by $b_{90}$, $a_{\rm fs,ij}$, or $R_1$, hence why we employ $\chi_1$ as the primary independent variable in all of our experiments.} References to \textit{hard}, \textit{semi-soft}, and \textit{wide} binaries correspond to binaries with $a_{\rm ij}<a_{\rm fs, ij}$, $a_{\rm ij}<10 \, a_{\rm fs, ij}$, and $a_{\rm ij}>R_1$, respectively. Accordingly, this means 3BBF probabilities for semi-soft and wide binaries include the contribution from hard binaries, but that contribution is negligible since 3BBF rates decrease sharply with binary hardness.

Throughout the entirety of our results, we find no evidence of any 3UB encounter or 3BBF event exhibiting resonant behavior in any regime. Here, resonance is the process by which a temporary bound state forms containing \textit{all} of the scattering bodies, usually characterized by a long, chaotic orbital dance. We know that none of our 3UB encounters feature resonance because at no point in any of our experiments do bound hierarchies form or dissolve after $\Delta t=2 t_{\rm e}$ has elapsed since starting the trajectory integration. In other words, if two bodies are bound to each other by time $2 t_{\rm e}$, they remain bound for all time ${>}2 t_{\rm e}$. Additionally, every single animation of 3BBF encounters we have produced in every regime (e.g., hard, soft, unequal-mass, etc.) are distinctively perturbative encounters, characterized by up to two slingshots shared by the three interacting bodies. The lack of resonance in our 3UB experiments is fully consistent with the understanding that resonant interactions are strongly disfavored, if not impossible, when the total energy in the center-of-mass frame of a three-body system is positive \citep{Heggie_2003, Binney_2008}.

\begin{figure}
\includegraphics[width=\columnwidth]{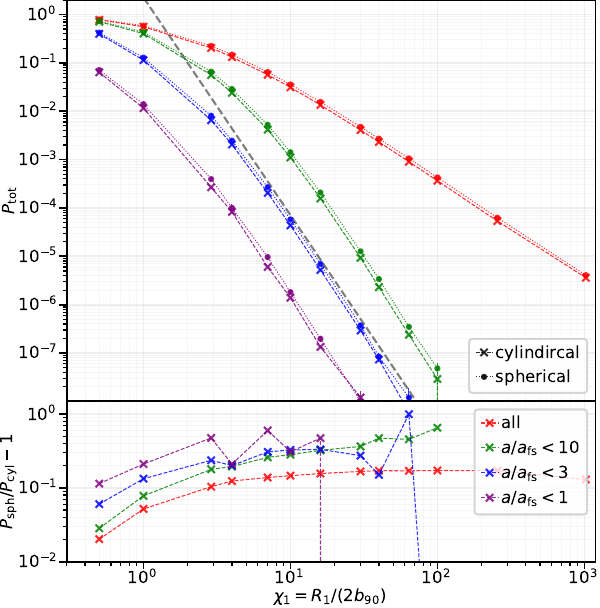}
\caption{Equal-mass 3BBF probabilities colored by minimum hardness with comparisons between the spherical and cylindrical (\AH{}) initial condition algorithms. \textit{Top Panel}: 3BBF probability, $P_{\rm ij}$, versus dimensionless impact parameter, $\chi_1$; all bodies are drawn from an identical Maxwellian velocity distribution. \dva{The $95\%$ confidence intervals are included as error bars for all points under the assumption 3BBF is a Poissonian process. However, these error bars are difficult to see for any probability ${\gtrsim}10^{-7}$ due to the large sample size ($10^9$ scatterings per value of $\chi_1$).}  As expected, $P_{\rm ij} \propto \chi_1^{-2}$ in the soft-binary limit (large $\chi_1$), in accordance with the scaling identified by \AH{}. Both algorithms produce a $P_{\rm ij} \propto \chi_1^{-4.5}$ decay for binaries with $a_{\rm ij}\lesssim R_1$ (the fit is the gray dashed line), as opposed to the $\chi_1^{-5}$ decay expected from standard analytic estimates; the discrepancy is likely due to the complicated nature of gravitational focusing in 3UB (see text). \textit{Bottom Panel}: Scatter plot quantifying the difference in probability between the two algorithms. We find that the harder the binary formed the greater the boost spherical sampling provides to 3BBF, up to ${\sim}30\%$ for hard binaries (purple).}
\label{fig:Pijvschi1_AHcompare}
\end{figure}

\begin{figure}
\includegraphics[width=\columnwidth]{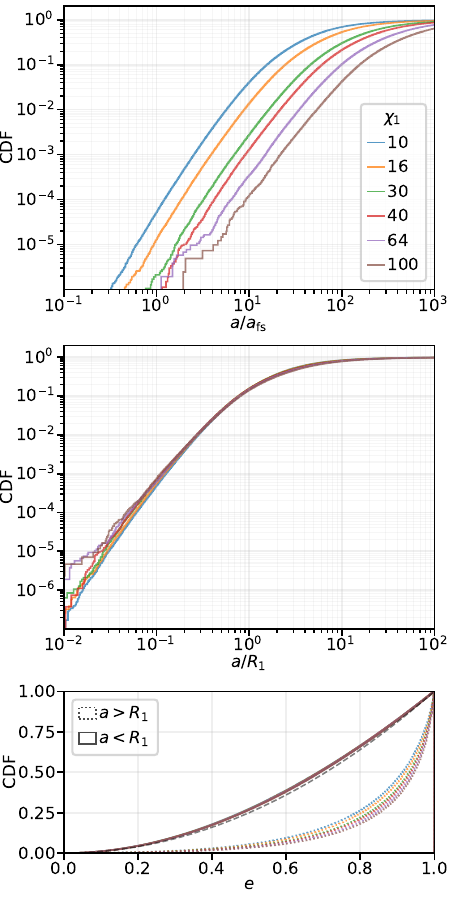}
\caption{\textit{Top Panel}: CDF of binary semi-major axis $a$ normalized by the fast-slow boundary $a_{\rm fs}$ during equal point mass 3BBF. Each curve represents an experiment with a different value of ${\chi_1 = R_1/(2 b_{90})}$. \textit{Middle Panel}: Identical to top panel, except the CDF is normalized instead by the radius of the encounter volume, $R_1$. Binaries formed with \dva{$a\lesssim R_1$} account for ${\approx}15\%$ of pairings. Independent of $\chi_1$, the cumulative semi-major axis distribution scales approximately as $C(a) \sim a^3$. \textit{Bottom Panel}: The corresponding CDFs for binary eccentricity $e$. Irrespective of $\chi_1$, the $e$ distribution for binaries with $a<R_1$ (overlapping solid curves) is nearly thermal, $C(e) \approx e^2$, while the $e$ distribution including all binaries (overwhelmingly soft binaries; dashed curves) is super-thermal, ${C(e) \approx e^\beta}$ with ${\beta>2}$. Thus, the widest binaries formed via 3BBF are well-described by super-thermal eccentricity distributions.}
\label{fig:smaecc_R1}
\end{figure}

\begin{figure}
\includegraphics[width=\columnwidth]{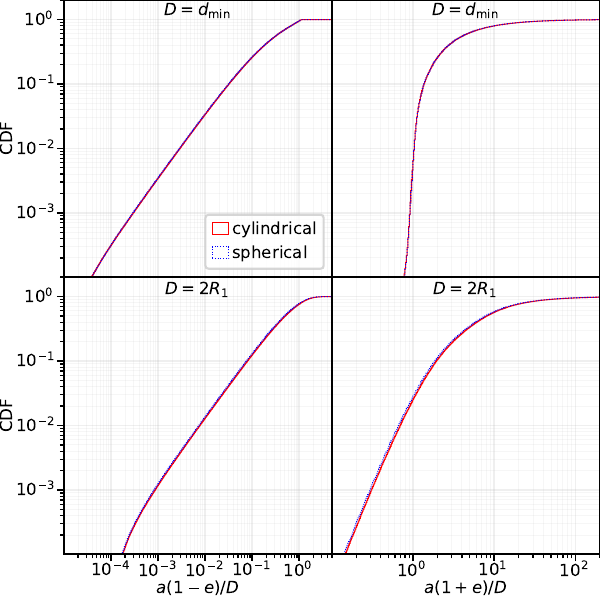}
\caption{CDFs of pericenter and apocenter distances normalized by (\textit{top}) the diameter $d_{\rm min}$ of the smallest sphere containing all three bodies at any time during the 3BBF interaction and (\textit{bottom}) the diameter of the interaction volume, $2 R_1$. These CDFs include every 3BBF in equal-point-mass 3UB interactions with $\chi_1=10$. Colors highlight the algorithm employed, the original cylindrical algorithm (solid red) from \AH{} versus our corrective spherical algorithm (dotted blue). In both cases, the minimum apoapse and maximum periapse distances are very nearly $d_{\rm min}$. This shows that $d_{\rm min}$ dictates the binary's orbital elements and implies that the fundamental physics of 3BBF is independent of the initial condition algorithm, so long as $R_2>R_1\gg 2 b_{90}$.}
\label{fig:r_dmin_hist}
\end{figure}

\subsection{Equal Point Masses}\label{sec:equalpointmasses}

In Figure~\ref{fig:Pijvschi1_AHcompare}, we show the 3BBF probability for the case of equal point masses as a function of the dimensionless impact parameter $\chi_1$. Colors distinguish binaries formed of different hardness while the point styles compare the outcome using the cylindrical \AH{} algorithm (crosses) versus our spherical correction to their algorithm (points). Spherical sampling boosts the 3BBF probability at all scales by ${\sim}10\%$--$50\%$ in comparison to \AH{}'s cylindrical sampling method. For both sampling methods, the 3BBF probability for binaries with $a_{\rm ij}<R_1$ and $\chi_1\gg 1$ scales as $P_{\rm ij}\propto \chi_1^{-4.5}$. This is slightly shallower than the $\chi_1^{-5}$ scaling one would expect from equating the 3BBF rate to the 3UB encounter rate \citep[e.g.,][]{Goodman_1993,Ivanova_2005,Ivanova_2010,Binney_2008,Morscher_2013,Morscher_2015}. We discuss the implications of this shallower $\chi_1$ dependency in Section~\ref{sec:hard_binary_formation}.

Binaries from 3BBF also exhibit several nearly geometrically scale-free properties when scaled to $R_1$. In the \dva{center} panel of Figure~\ref{fig:smaecc_R1}, we show the cumulative distribution for binary semi-major axes (SMA) from 3BBF, normalized by the radius of the interaction volume, $R_1$. In the bottom panel, we display the eccentricity distribution of binaries with $a>R_1$ and binaries with $a<R_1$. In both panels, the color of the distribution denotes the value of $\chi_1$ used in the scattering experiment. We find that both the SMA and eccentricity distributions do not depend on $\chi_1$ in that they are nearly independent of the size of the interaction volume (geometrically scale-free). Specifically, the binaries with $a<R_1$ are well-described by the thermal eccentricity distribution (dashed black; ${f(e)=2 e,\, C(e)=e^2}$) while binaries overall (dominated by those with $a>R_1$) have super-thermal eccentricity.

It may appear counter-intuitive that 85\% of the binaries have an SMA larger than the radius of the interaction volume, $a>R_1$, and super-thermal eccentricities. However, there is no upper limit on binary SMA, and $R_1$ is typically larger than $d_{\rm min}$ (the diameter of the smallest sphere containing all three bodies at any point during the interaction), fixing the maximum possible angular momentum available to 3BBF. We find that $d_{\rm min}$ naturally sets both the maximum periapse, $a(1-e)<d_{\rm min}$, and the minimum apoapsis, $a(1+e)>d_{\rm min}$. Independent of algorithm and binary hardness, our experiments show that the former inequality is satisfied in $100\%$ of 3BBF, while the latter is satisfied in ${>}99\%$ of 3BBF  (Figure~\ref{fig:r_dmin_hist}). So, to satisfy $a(1-e)<d_{\rm min}$, binaries from 3BBF with large SMA ($a > d_{\rm min}$) must have high eccentricity ($\langle e\rangle \sim 1$; a super-thermal distribution).

\begin{figure*}
\includegraphics[width=\textwidth]{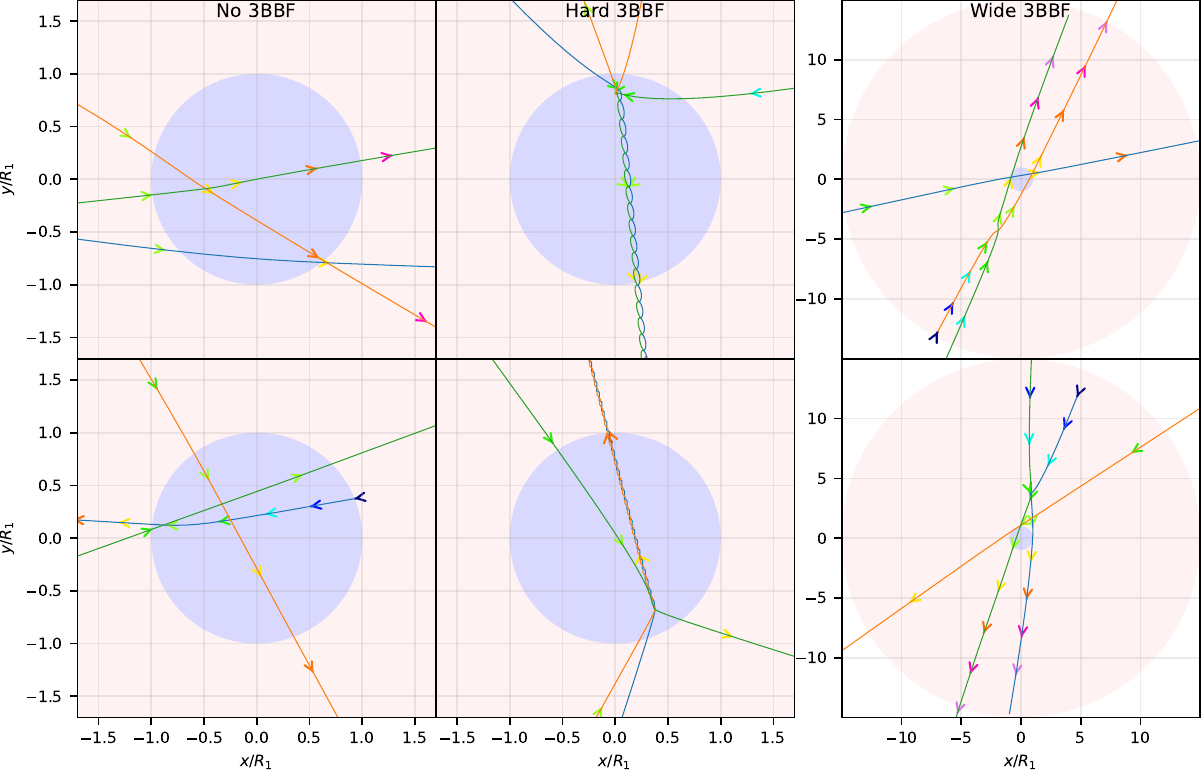}
\caption{\dva{A 2D projection of six different equal-mass 3UB realizations. Each column corresponds to scatterings that result in (i) \textbf{no}, (ii) \textbf{hard} (here, $a \ll R_1$), or (iii) \textbf{wide} ($a > R_1$) 3BBF. The blue shaded region is the interaction volume of the initial condition algorithm (with radius $R_1$) and the red shaded region is defined by $R_2$; see Section~\ref{sec:IC_Alg} for details. Line color represents the trajectory of an individual body while color-coded arrows show the direction bodies are moving at uniformly selected color-coded time steps in the simulation. The larger the spacing between arrows, the faster the body is moving. No resonant encounters occur here or in any 3UB simulation made yet. }}
\label{fig:3UBtrajectories}
\end{figure*}

While we find wide binaries ($a>R_1$) from 3BBF are super-thermal, our results unambiguously confirm that binaries with $a<R_1$, encompassing all hard binaries, are born with thermal eccentricities. This is a classic prediction for both hard and soft binaries under the assumption they have undergone many successive encounters within their environment \citep{Jeans_1919, Heggie_1975}. \dva{Sample 3UB trajectories resulting in hard 3BBF, wide 3BBF, and no 3BBF (a simple flyby of three single bodies), are displayed in Figure~\ref{fig:3UBtrajectories}. The figure shows that 3BBF occurs as a single, non-resonant event, implying that the thermal eccentricity distribution may be more fundamental to binary formation than previously assumed, independent of detailed balance or the need for an individual binary to  undergo many successive encounters.}

\subsubsection{Choice of $\chi_1$}\label{sec:choice_of_chi1}
Having examined how the 3BBF probability depends on $\chi_1$ in the case of equal point masses, it is natural to wonder what is a proper choice of $\chi_1$. Setting $\chi_1$ high enough that $R_1$ exceeds the interparticle distance in the host environment would clearly violate the assumed isolation of the interaction from its surroundings. Yet simply anchoring $R_1$ to the interparticle distance is not computationally optimal; since the 3BBF probability drops steeply with increasing $\chi_1$, choosing too large a $\chi_1$ needlessly inflates the number of scattering experiments required to achieve a robust sample of binaries. However, $\chi_1$ cannot be made arbitrarily small without biasing the properties of the binaries that are formed.

In the limit $\chi_1\rightarrow0$, the initial positions and velocities of all bodies are focused toward a point because $R_2 = 15 R_1$ in the \AH{} method. This minimizes the angular momentum in the global reference frame and causes the 3BBF probability to saturate (no longer follow a simple power law scaling with $\chi_1$). By initializing the bodies deeper within their mutual potential wells, this limit also results in a total initial energy much nearer to zero than in a scattering experiment conducted at higher $\chi_1$. Since the 3UB problem requires total positive energy, shrinking $\chi_1$ too far would bias the initial conditions to be just barely unbound. From an algorithmic perspective, the vast majority of randomly generated initial conditions in this limit would sample at least two bound bodies, and have to be thrown out. This artificially truncates the energy--angular momentum parameter space that would otherwise be obtained naturally from the isotropically sampled Maxwellian velocity distributions.

\dva{A key limiting factor in choosing $\chi_1$ is the average interparticle distance, $\langle r \rangle$, of the 3UB interaction's host environment. Specifically, to satisfy the assumption of an isolated 3UB encounter, then $R_1$ must be ${<} \langle r \rangle/2$.\footnote{To ensure an isolated encounter, we assign the extra factor of two here so that $\langle r \rangle$ is the maximal possible \textit{diameter} of the interaction volume.} We can form a qualitative picture by first writing $\langle r \rangle$ in terms of fundamental quantities of the Plummer model, 
\begin{equation}
    \begin{aligned}
        \langle r \rangle = \frac{G M_{\rm P}}{6} \frac{\sigma_{\rm c}^{4/3}}{\sigma^{10/3}} N^{-1/3},  
    \end{aligned}
\end{equation}
where $M_{\rm P}$ is the total mass of the Plummer cluster, $\sigma_{\rm c} = \sqrt{\frac{G M}{6 \, b_{\rm P}}}$ is the Plummer core velocity dispersion, $b_{\rm P}$ is the Plummer scale length, $\sigma$ is the local velocity dispersion, and $N\approx M_{\rm P}/\langle m\rangle$ is the number of bodies in the cluster. In normalizing $\langle r \rangle$ by $4 b_{90}$, we can probe how it relates to the $\chi_1$ scaling employed in this work. To satisfy $R_1 < \langle r \rangle/2$, we must set $\chi_1 < \chi_{\rm r}$, where
\begin{equation}
\begin{aligned}
    \chi_{\rm r} = \frac{\langle r \rangle} {4 b_{90}} &= \frac{1}{8} N^{2/3} \left(\frac{\sigma_{\rm c}}{\sigma}\right)^{4/3}\\
    &= \frac{1}{8} \left[N \left(1 + \frac{r^2}{b_{\rm P}^2} \right) \right]^{2/3},
    \end{aligned}
\end{equation}
with $r$ as the radial distance from the center of the Plummer cluster. If we limit our investigations to the core of a cluster, we find that $\chi_{\rm r} \approx 0.125 \, N^{2/3}$ and that clusters consisting of $N=\{10^3, 10^4, 10^5, 10^6, 10^7\}$ bodies have a dimensionless $\chi_{\rm r} \approx \{13, 60, 270, 1250, 5800\}$, respectively. These choices of $N$ span the typical range for open clusters to dense nuclear star clusters.}

To choose an appropriate $\chi_1$ for the rest of our analysis---and thereby enable more thorough examination of other important considerations for 3BBF---we take guidance from the above estimates surrounding $\chi_{\rm r}$ and Figures~\ref{fig:Pijvschi1_AHcompare} and \ref{fig:smaecc_R1}. In particular, the scalings of the 3BBF probabilities on $\chi_1$ asymptotically settle into simple power laws at $\chi_1\gtrsim 10$, \dva{enabling straightforward extrapolation in dynamically active environments that can sustain larger values of $\chi_{\rm r}>\chi_1>10$.} Beyond this point, the properties of the binaries, such as semi-major axis (normalized by $R_1$) and eccentricity, no longer depend on $\chi_1$. We find this also holds for unequal mass simulations. 

\dva{Therefore, we choose to use a default value of $\chi_1=10$ in all our following analysis, unless noted otherwise. For this choice of $\chi_1$, ${\approx}0.7\%$ of all 3UB initial conditions result in at least two of the three bodies already being energetically bound to each other. These 3UB initial conditions are rejected since we are interested in the formation of \textit{new} binaries. Although this rejection rate is larger than the typical 3BBF probability at $\chi_1=10$, our tests indicate that including scatterings where bodies are allowed to be bound at initialization does not significantly alter the soft or hard 3BBF probabilities in the equal- and unequal-mass cases. It follows that 3UB initial conditions in which two bodies are initiated especially close to one another are not a significant source of 3BBF. Our choice of $\chi_1=10$ therefore balances both accuracy and computational efficiency.}

\subsubsection{Hard Binary Formation}\label{sec:hard_binary_formation}

We now estimate the hard 3BBF rate in the equal point mass limit using the results of our analytically derived encounter rate---Equation~(\ref{eq:gammamax_ijk})---and numerically determined formation probabilities (Figure~\ref{fig:Pijvschi1_AHcompare}). We assert that the general solution to a numerically determined volumetric 3BBF rate is of the form
\begin{equation}
    \tilde{\Gamma}^{\rm F} \left(<a,\chi_1 \right) = P \left(<a, \chi_1\right) \tilde{\Gamma}_{\rm 3B}\left(\chi_1 \right),
\end{equation}
where $\tilde{\Gamma}_{\rm 3B}$ is the 3UB volumetric encounter rate and the 3BBF rate per 3UB encounter, $P \left(<a,\chi_1\right)$, is extracted from Figure~\ref{fig:Pijvschi1_AHcompare}. 

\dva{Simplified for the case of identical masses with velocities drawn from a single shared velocity distribution with dispersion $\sigma$, the volumetric encounter rate is
\begin{equation}\label{eq:equalmassenc}
    \tilde{\Gamma}_{\rm 3B} = \frac{2^{5/2} \pi^{3/2}}{3} n^3 R_1^5 \sigma.
\end{equation}
}
The cumulative distribution of hard-binary SMA has the form $P(<a) \propto \left(a/a_{\rm fs}\right)^3$ when $a<R_1$. In Figure~\ref{fig:Pijvschi1_AHcompare}, the cumulative probability of forming binaries with ${a/a_{\rm fs}\leq 1}$ is 

\begin{equation}
P(<a,\chi_1) \approx1.85 \times 10^{-6} \left(\frac{\chi_1}{10}\right)^{-4.5} \left(\frac{a}{a_{\rm fs}}\right)^3.
\end{equation}
Combining this probability with our encounter rate, the hard 3BBF rate for equal masses is 
\dva{
\begin{equation}\label{eq:estimated_formation_rate}
    \tilde{\Gamma}^{\rm F}\left(<a, \chi_1\right) 
    \approx 0.081 \, \frac{G^5 m^5 n^3}{\sigma^9} \left(\frac{a}{a_{\rm fs}}\right)^3 \chi_1^{1/2}.
\end{equation}
Note that the $\chi_1^{1/2}$ dependency in Equation~(\ref{eq:estimated_formation_rate}) highlights a unique environmental constraint on 3BBF rates not accounted for in any previous works. While the formation rate appears to be divergent---$\tilde{\Gamma}^{\rm F} \rightarrow \infty$ as $\chi_1 \rightarrow \infty$---this is contingent on the existence of an environment with an infinite average interparticle distance, a nonphysical consideration. Thus, to properly estimate a local 3BBF rate, a careful determination of the largest possible $\chi_1<\chi_{\rm r}$ must be used in future (semi-)analytic prescriptions. Recently, \citet{Ginat_2024_3BBF} corroborated our $\chi_1^{-4.5}$ probability scaling, finding an identical scaling by evaluating their analytic framework with their least stringent limiting condition on the 3UB interaction volume.}

\begin{figure}
\begin{center}
\includegraphics[width=\columnwidth]{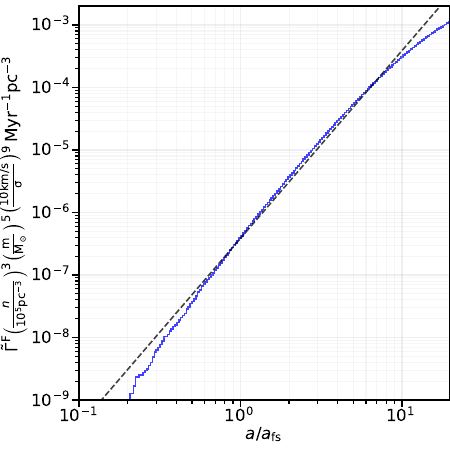}
\caption{The estimated cumulative 3BBF rate for the interaction of three equal-mass bodies sampled from identical Maxwellian velocity distributions. Here, we show only the scattering experiments with dimensionless impact parameter ${\chi_1=10}$. The blue curve is a CDF of semi-major axis ($a/a_{\rm fs}$) scaled by Equation~(\ref{eq:3BBF_vol_rate}) and the black dashed line is a curve fit defined by Equation~(\ref{eq:estimated_formation_rate}). The physical cross section for 3BBF increases with mass and decreases with speed since $R_1\propto m/\sigma^2$. Our rate is within $15\%$ of the rate predicted by \citet{Goodman_1993}; see text near Equation~(\ref{eq:estimated_formation_rate}).}
\label{fig:formationrate_vs_sma}
\end{center}
\end{figure}

If we consider $\chi_1=10$ as a test case and redefine our hardness criteria in terms of the hard-soft boundary as defined by \citet{Goodman_1993}, $a_{\rm hs} = \frac{G m_1 m_2}{2 m \sigma^2}$, we may substitute $a_{\rm fs}\rightarrow \frac{2}{3} a_{\rm hs}$. The formation rate, cumulative in $a$ such that it includes all SMA ${<}a$, then becomes ${\tilde{\Gamma}^{\rm F}\left({<}a\right) \approx 0.86 \, \frac{G^5 m^5 n^3}{\sigma^9} \left(\frac{a}{a_{\rm hs}}\right)^3}$. The coefficient of $0.86$ is a close match to the prediction of $0.75$ from \citet{Goodman_1993} using detailed balance (and integrating their Equation~2.6 from $x/(m\sigma^2):[1\rightarrow\infty]$), sans the SMA and $\chi_1$ scalings we have identified. This semi-analytic expression for the 3BBF rate is displayed as a function of SMA in Figure~\ref{fig:formationrate_vs_sma} and over-plotted atop the numerical results for equal masses and $\chi_1=10$.

\begin{figure}
\includegraphics[width=\columnwidth]{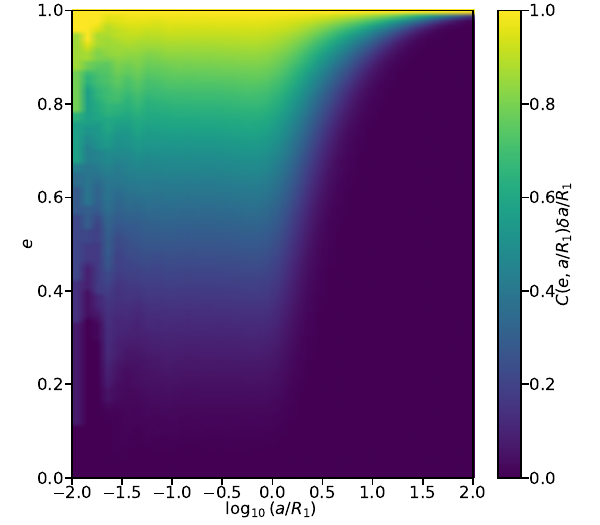}
\caption{Heat map of binary eccentricity versus SMA (normalized by $R_1$, the radius of the interaction volume) across all 3BBF scatterings with $\chi_1\geq10$. Each (vertical) bin in SMA is normalized by the cumulative probability of binary formation in that bin $C(a/R_1)$. Eccentricity is thermal and roughly independent of SMA for $a/R_1\lesssim1$, but skews increasingly super-thermal with increasing $a/R_1$ for $a/R_1\gtrsim 1$ (${\approx}98\%$ of binaries).}
\label{fig:smaecc_R1_heatmap}
\end{figure}

\subsubsection{Wide Binary Formation}\label{sec:widebinaries}
Soft binaries may also be described by the same scaling relations shown in the previous section so long as the parameter space of binary properties are restricted to $a<R_1$. The opposite limit---$a>R_1$---hosts properties exclusively describing the widest binaries that may form through 3BBF. These binaries follow an entirely different binary formation probability curve ($P\propto \chi_1^{-2}$) and their SMA and eccentricity distributions are not described by a simple power law, unlike in the hard binary limit. Given that binary SMA and eccentricity distributions are nearly independent of $\chi_1$ for $\chi_1 \in [10,100]$ (Figure~\ref{fig:smaecc_R1}), we may combine the scattering experiments spanning this interval to examine these binary properties at higher resolution. As a reminder, $\chi_1$ represents a numerical factor chosen for convenience, with a maximal value dependent on environmental properties (e.g., the local average inter-particle separation).

Figure~\ref{fig:smaecc_R1_heatmap} is a two-dimensional heat map of the resulting SMA--eccentricity space from this combined dataset. The underlying density is normalized in each (vertical) SMA bin to aid visualization of the eccentricity distribution for any given SMA. Figure~\ref{fig:smaecc_R1_heatmap} shows that the eccentricity distribution is roughly independent of SMA for $a/R_1<1$ (with some fluctuation attributable to low resolution at low SMA). However, for $a/R_1>1$, the binaries formed from 3BBF skew increasingly eccentric (super-thermal) as SMA increases, with the absolute softest binaries formed with a fixed interaction volume being exclusively super-thermal. 

Following the same calculation as our hard binary formation rate from the previous section, the formation rate for super-thermal wide binaries crucially depends on the size of the interaction volume, $R_1$. Additionally, a power-law fit to the red curve in Figure~\ref{fig:Pijvschi1_AHcompare} for $\chi_1>10$ yields a functional form for the overall 3BBF probability of $P\approx4.2\chi_1^{-2}$. Since ${\approx}85\%$ of those binaries are wide (have $a>R_1$ in Figure~\ref{fig:smaecc_R1}), then the probability of forming a wide binary from 3BBF in the equal-mass limit at $\chi_1>10$ is roughly
\begin{equation}
    P(a>R_1) \approx 3.6 \chi_1^{-2} = 14.4 \frac{b_{90}^2}{R_1^2} = 1.6 \frac{G^2 m^2}{\sigma^4 R_1^2}.
\end{equation} 
Combining this probability with equation~\ref{eq:equalmassenc}, the volumetric 3BBF rate for (super-thermal) wide binaries in the equal-mass limit is
\begin{equation}
    \tilde{\Gamma}^{\rm F}(a>R_1) \approx 16.8 \frac{G^2 m^2 n^3}{\sigma^3} R_1^3.
\end{equation}
Additionally, ${>}50\%$ of wide binaries have an SMA between $1<a/R_1<5$ (see Figure~\ref{fig:smaecc_R1}). 

The ability to realize a super-thermal, wide binary distribution is contingent on the dynamical properties of environments hosting 3BBF. Dynamically active and well-populated environments (e.g., star cluster cores) may enable many successful 3BBF events, but reasonably long-lived binaries form hard, with SMA smaller than the average inter-particle separation (i.e., $a<R_1<\langle r\rangle$). Newly formed binaries with SMAs larger than the average inter-particle separation (and super-thermal eccentricities) are highly unlikely to persist within dense environments. In effect, binaries born within a central, dynamically active region should be thermal, independent of binary binding energy.


Yet, open clusters, star cluster halos, and stellar streams may not be so prohibitive \citep{Pen_2021}.  The isotropically distributed recoil velocity experienced by all new binaries in 3BBF may quickly dissociate them from loose environments with low escape velocity, enabling the formation of binaries wider than the $\langle r_{\rm sep}\rangle$ of their original host. Thus, in contrast with other dynamical methods which do not explicitly investigate 3UB interactions \citep{Hamilton_2023, Xu_2023}, the 3BBF mechanism may dynamically populate the super-thermal wide binaries observed by \textit{Gaia} in the galactic field \citep{Tokovinin_2020,Hwang_2022}.

\begin{figure*}
\begin{center}
\includegraphics[width=\textwidth]{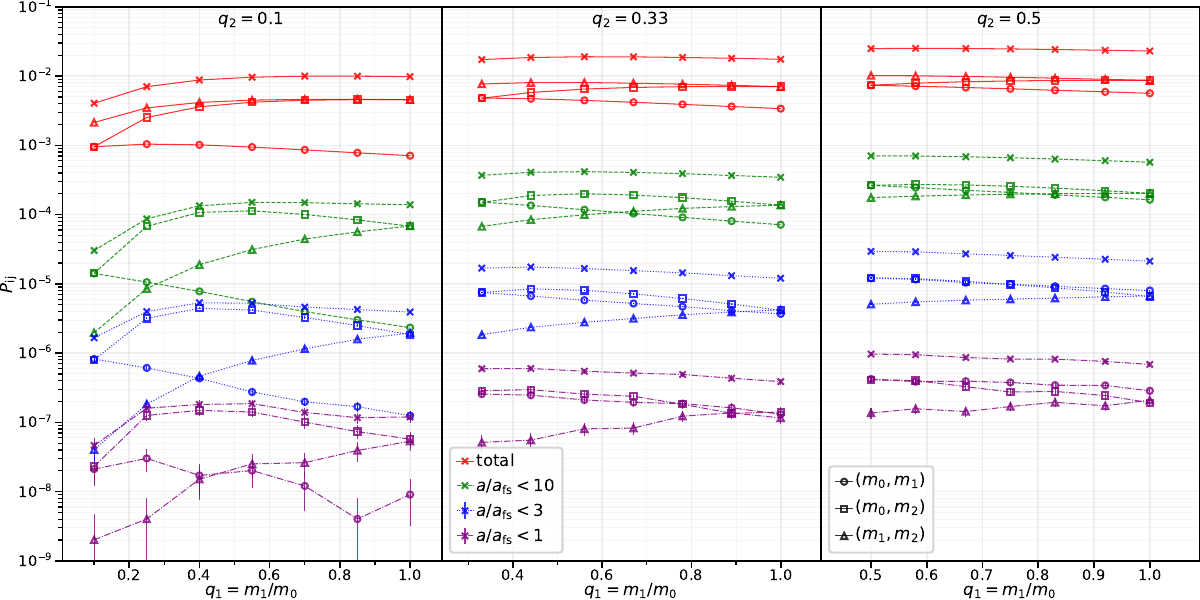}
\caption{Probability of binary formation, $P_{\rm ij}$, versus mass ratio of the second most massive body, ${q_1 = m_1/m_0}$. Mass ratios, ${q_{\rm i} = m_{\rm i}/m_0}$, are labeled in descending order by mass, with $q_0=1$ and ${q_2\leq q_1 \leq 1}$. The $95\%$ confidence intervals are included as error bars for all points under the assumption 3BBF is a Poissonian process. However, these error bars are difficult to see for any probability ${\gtrsim}10^{-7}$ due to the large sample size ($10^9$ scatterings per value of $\chi_1$). Accounting for hardness, we find that the pairing of the two most massive bodies is the least likely across the majority of the parameter space; the two least-massive bodies are the most likely pairing when allowing for binaries of arbitrarily large SMA.  Each panel (column) shows $P_{\rm ij}$ for a different fixed ${q_2 = m_2/m_0}$. Color distinguishes binaries with the different minimum hardnesses specified in the legend while shapes indicate the three possible pairings of bodies. For most unequal mass ratios, new binaries typically contain the \textit{least} massive body. The total 3BBF probability (cross-hatched curves for each color) increases with $q_2$ (i.e., as the mass ratios approach unity), regardless of binary hardness. See Table~\ref{table:q} for a numeric list of these probabilities and their counting uncertainties.}
\label{fig:sigvseta_q}
\end{center}
\end{figure*}

\begin{figure*}
\begin{center}
\includegraphics[width=.95\textwidth]{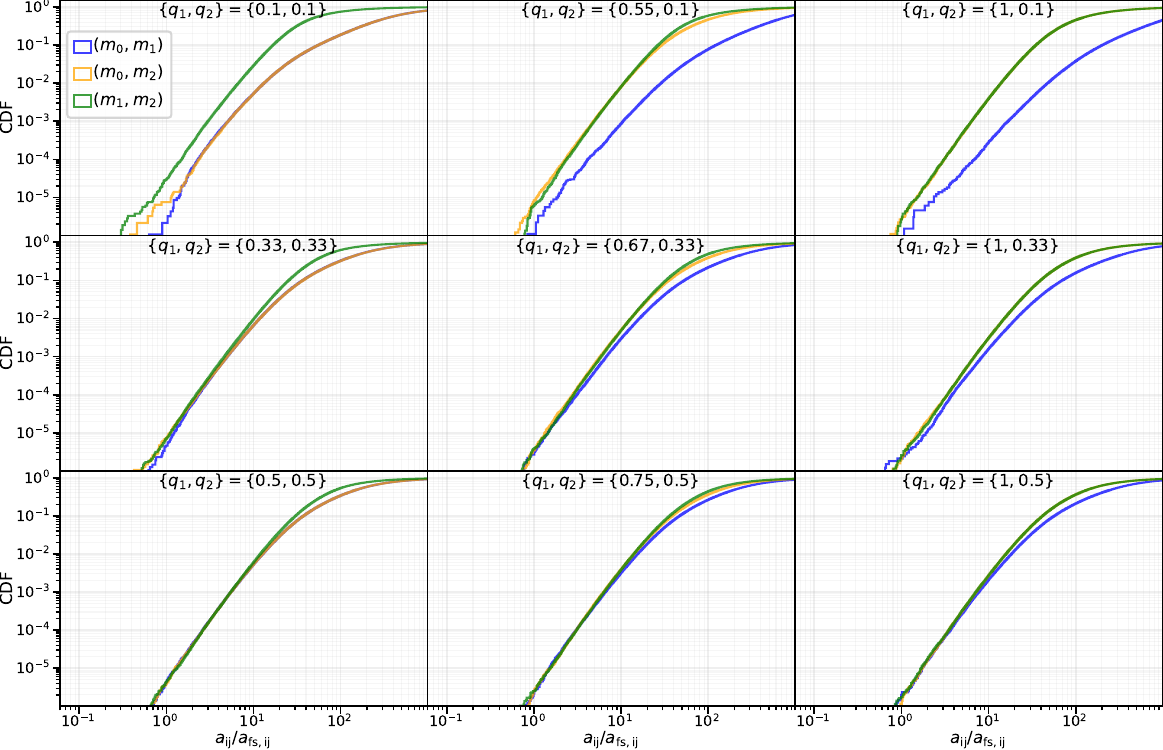}
\caption{Binary SMA CDFs across various mass ratios with fixed velocity dispersion and $\chi_1=10$. Mass ratios, ${q_{\rm i} = m_{\rm i}/m_0}$ with $m_0\geq m_1\geq m_2$, are seen at the top of each subplot. Color specifies the pairing $\{q_{\rm i}, q_{\rm j}\}$ of each distribution. For $a_{\rm ij} < R_1$ (here, $R_1 = 2 \chi_1 a_{\rm fs,01}$), the SMA distributions exhibit a simple power-law relation similar to the equal-mass case, but feature non-trivial deviations in the wide-binary limit ($a_{\rm ij} > R_1$).}
\label{fig:sma_q}
\end{center}
\end{figure*}

\begin{figure*}
\begin{center}
\includegraphics[width=.95\textwidth]{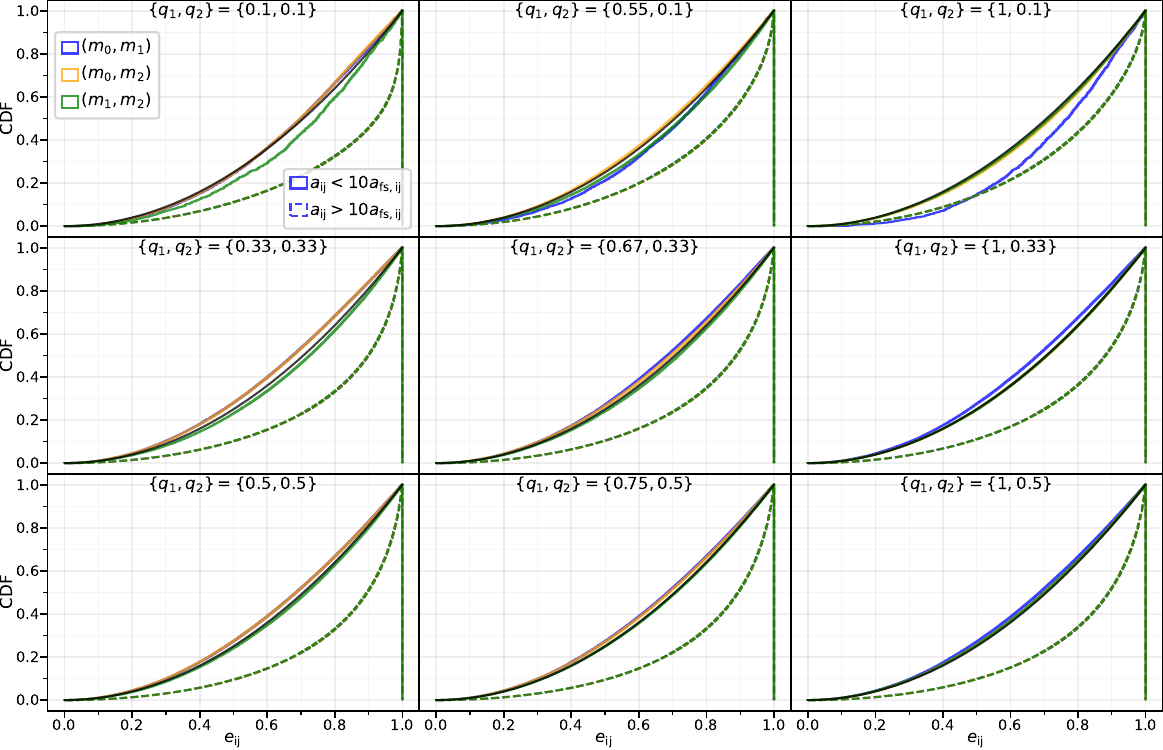}
\caption{Binary eccentricity CDFs across various mass ratios with fixed velocity dispersion and $\chi_1=10$. Mass ratios, ${q_{\rm i} = m_{\rm i}/m_0}$ with $m_0\geq m_1\geq m_2$, are seen at the top of each subplot. Color specifies which two of the three bodies end up in the binary, numbered in decreasing order of their mass (0,1,2). The \{solid, dashed\} line style corresponds to binaries with SMA \{greater, less\} than $10 a_{\rm fs,ij}$. Wide binaries, those with $a_{\rm ij}\gtrsim10 a_{\rm fs,ij}$, exhibit identical super-thermal eccentricity distributions across mass-pairings for fixed masses.}
\label{fig:ecc_q}
\end{center}
\end{figure*}

\subsection{Unequal Point Masses}\label{sec:unequal_mass}
We now report our findings on the first investigation of 3BBF for the case of unequal point masses. \dva{Just as in our equal-mass investigation, each body has an initial velocity randomly drawn from a single shared Maxwellian velocity distribution, with position and velocity unit vectors assigned according to our algorithm (see Section~\ref{sec:IC_Alg}). In total, this dataset contains exactly $2.1\times10^{10}$ simulations, or $10^9$ scatterings per mass ratio combination.} 

Figure~\ref{fig:sigvseta_q} summarizes our results. The top row is the 3BBF probability, $P_{\rm ij}$, as a function of mass ratio, $q_{\rm i}$. We explore the unequal-mass 3UB parameter space by transforming masses as: ${(m_0,m_1,m_2)\rightarrow(q_0, q_1, q_2)}$ with ${q_{\rm i}=m_{\rm i}/m_0}$, ${{q_2 \leq q_1 \leq q_0}}$, and fixing $q_0=1$. Colors denote the minimum hardness ($a_{\rm ij}/a_{\rm fs,ij}<x$) of the binaries considered in each SMA bin while shapes separate binaries by the mass pairing $\{i,j\}$. The dimensionless impact parameter is fixed to $\chi_1=10$ for all experiments as we find that it balances resolution, efficiency, and accuracy throughout the entire parameter space (see Section~\ref{sec:equalpointmasses}).

A key assumption of 3BBF prescriptions in star cluster modeling codes such as \texttt{CMC} \citep[e.g.,][]{Morscher_2013,Morscher_2015} is that the two most massive bodies $(q_0, q_1)$ in a 3UB encounter are the most likely to pair into a binary. Our findings unambiguously reject this assumption. Including binaries of any size, it is instead the two least massive bodies $(q_1, q_2)$ which are the most likely to pair. The most plus least massive bodies $(q_0, q_2)$ are the second most likely to pair generally, but the most likely in the hard binary limit. The pairing of $(q_0, q_1)$ is the least likely 3BBF end-state independent of hardness, becoming orders-of-magnitude less likely as $q_2$ approaches the test particle limit.

Naively, the tidal effect experienced between two bodies within the gravitational field of a third scales as $\left(\frac{m_{\rm k}} {m_{\rm i} + m_{\rm j}}\right)^{1/3}$, where $m_{\rm k}$ is the mass of the perturbing body. It follows that it is significantly easier for a more massive particle to perturb two low mass bodies than for a low mass body to perturb two high mass bodies. Stated differently, it may be easier to change the energy/momentum of a less-massive body (less inertia), making it easier to extract two-body energy if the leftover single is more massive. That said, a more thorough explanation for the process of unequal mass 3BBF is beyond the scope of the current work and will be investigated in detail in the future. 

The differences in the pairing probabilities become increasingly subtle the closer in mass the three bodies are. For most of the explored parameter space, the pairing of $(q_0, q_2)$ is the most probable hard binary pairing, with the most massive bodies $(q_0, q_1)$ and least massive bodies $(q_1, q_2)$ swapping prevalence as $q_1\rightarrow1$. As mass ratios approach unity, hard $(q_0, q_1)$ pairings become more probable. Still, the pairing of the most massive bodies always comprises ${<}50\%$ of total pairings, independent of hardness.

Turning to binary SMA and eccentricity distributions for fixed mass ratios (Figures~\ref{fig:sma_q} and \ref{fig:ecc_q}), many of the tendencies occurring in equal-mass scattering are asymptotically emergent as binaries approach the fast/slow boundary. For the most extreme mass ratios $(q_2=0.1)$, $C(a_{\rm ij})\propto a_{\rm ij}^{2.5-3.0}$ and the SMA for the pairing of the most massive bodies $(q_0, q_1)$ scales as $a_{\rm 01}^{2.5}$. As $q_2\rightarrow1$, all SMA distributions tend towards $C(a_{\rm ij})\propto a_{\rm ij}^{3.0}$, as earlier identified in the case of equal masses. 

Eccentricity distributions for binaries with ${a_{\rm ij}/a_{\rm fs,ij}<10}$ follow an identical trend to what we identified with equal-mass encounters: they closely follow a thermal distribution. The only exception is the pairing of the two high-mass bodies, which yields a mildly super-thermal eccentricity distribution. Meanwhile, the eccentricity distributions for soft binaries are more extreme than in the case of equal-mass scattering. The soft pairing of $(q_0,q_1)$ is extremely super-thermal $(\langle e_{01}\rangle\gtrsim0.95)$ while the soft pairing of $(q_1,q_2)$ tends closer to a thermal distribution than in the equal-mass case.

\begin{figure*}
\begin{center}
\includegraphics[width=4in]{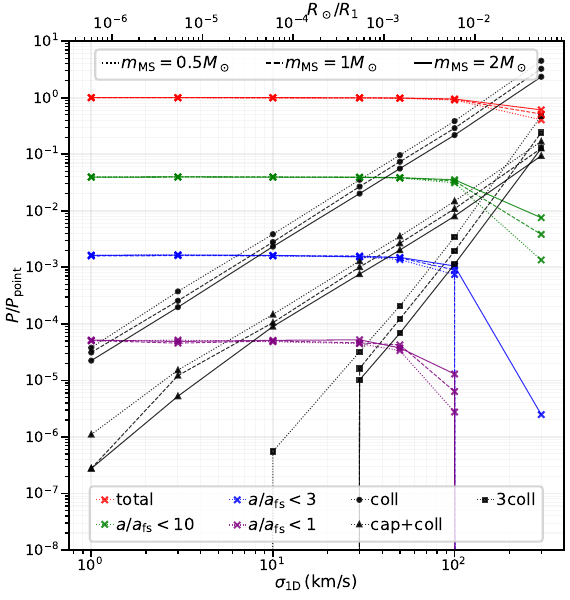}
\includegraphics[width=3in]{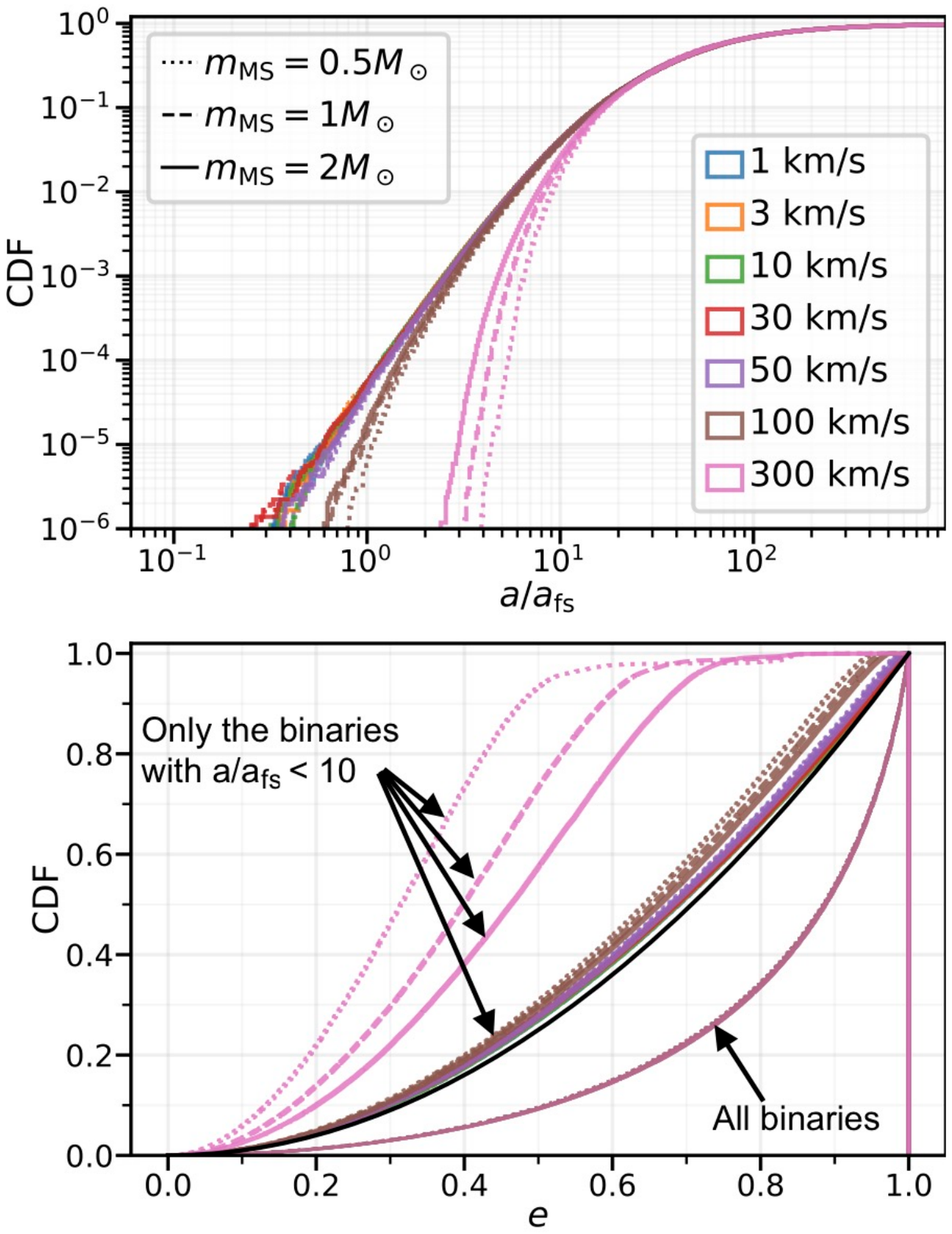}
\caption{
3BBF distributions from scattering finite-sized bodies with main-sequence (MS) star radii, varying the one-dimensional velocity dispersion $\sigma$ and stellar mass but keeping $\chi_1=10$ fixed.\\
\textit{Left}: Probability of 3BBF normalized by the total 3BBF probability (for $\chi_1=10$) in the earlier point mass scenario (red curve in Figure~\ref{fig:Pijvschi1_AHcompare}). Line style denotes the mass of the MS star, color again indicates binary hardness, while black indicates collision probabilities. The black labels are: \textbf{coll}---one collision, \textbf{cap+coll}---collision and capture of the third body, \textbf{3coll}---all three bodies collide during scattering.
For most (but especially higher) velocities, collision probabilities exceed 3BBF probabilities. Even so, collisional effects negligibly reduce the 3BBF probability for $\sigma\lesssim 30\,\rm{km \, s^{-1}}$.\\
\textit{Right}: Cumulative SMA and eccentricity distributions for each velocity dispersion. As reflected in the formation rates at left, the SMA distribution is roughly independent of $\sigma$ until $\sigma\gtrsim30\,\rm{km \, s^{-1}}$, where hard 3BBF becomes increasingly unlikely. Soft binaries are universally described by the same super-thermal eccentricity distribution (set of overlapping curves in the lower half of the eccentricity panel) independent of $\sigma$. Binaries with $a/a_{\rm fs}<10$ all closely follow a thermal eccentricity distribution (black curve) until the distributions become disrupted by the prevalence of collisions when $\sigma\gtrsim30\,\rm{km \, s^{-1}}$, quickly suppressing formation of hard, eccentric binaries.}
\label{fig:sigvseta_finitesize}
\end{center}
\end{figure*}

\section{Pairing Stars and Black Holes}\label{sec:finitesize_PN}

We now explore for the first time 3BBF between MS stars with masses $m_{\rm MS}/M_\odot: \{0.5, 1, 2\}$ and stellar mass BHs with masses $m_{\rm BH}/M\odot: \{20, 50, 100\}$. This is also the first study consider either relativistic or finite-size effects within the context of 3UB encounters. Additionally, we explore 3BBF between the encounter of: (i) two BHs and an MS star and (ii) two MS stars and a BH---mixed-species encounters that dominate 3BBF at most times in models of typical Milky Way globular clusters \citep[e.g.,][]{Weatherford_2023a}.

As in our point mass simulations, we fix the dimensionless impact parameter to be $\chi_1=10$ and draw the velocities of all bodies from a Maxwellian with a one-dimensional velocity dispersion, $\sigma$. In reality, species with such different masses will typically have different velocity distributions, but we leave nuanced exploration of this further complexity to future work to focus on the raw influence of mass ratio, along with relativistic and finite-size effects. Variation in $\sigma$ also varies the impact parameter, since ${R_1 \propto \sigma^{-2}}$, so it probes the strength of these effects in close passages. To show this impact clearly in our results, we therefore vary the velocities in the range $\sigma \in [1,300]\,\rm{km\,s}^{-1}$ for MS stars and $\sigma \in [3,3000]\,\rm{km\,s}^{-1}$ for BHs. Varying mass to probe relativistic and finite-size effects effects is also an option, but the typical mass distributions of evolved MS stars and stellar-mass BHs in dense star clusters each span a smaller range than the typical local mean-squared velocities of the various dynamically active environments they may inhabit  (e.g., open clusters to nuclear star clusters). Also, 3BBF rates scale more steeply with $\sigma$ than with mass.

To explore the finite-size effects, we assign radii to stellar bodies according the classic MS radius relation \citep{Demircan_1991},
\begin{equation}\label{eq:msradius}
    r_{\rm i} = (m_{\rm i}/M_\odot)^{3/5} R_\odot .
\end{equation}
To mitigate inaccuracies in the post-Newtonian (PN) approximation near the event horizon, BHs are assigned radii of $7$ times their Schwarzschild radius, 
\begin{equation}
    r_{\rm i} = 14 \frac{G m_{\rm i}}{c^2}.
\end{equation}
In all cases, collisions between stars are handled with the sticky sphere approximation, combining the masses (with no mass loss) once the surfaces of stellar bodies touch. \dva{The collison product is then placed instantaneously at the center of mass of the two former bodies with their center-of-mass velocity.} BH mergers are treated with the numerical relativity prescriptions of \citet{Lousto_2013, Healy_2018}. Collisions between BHs and MS stars are resolved by placing the BH at the center of mass of the star--BH pair with their mutual center-of-mass velocity and assume no accretion (the star is destroyed). Under all circumstances, PN terms up to PN3.5 are enabled during integration and all BHs are assumed initially non-spinning. Tidal physics is not included in this first exploration of non-point-mass 3UB interactions, but we plan to explicitly explore such effects in a later work.

\begin{figure*}
\begin{center}
\includegraphics[width=4in]{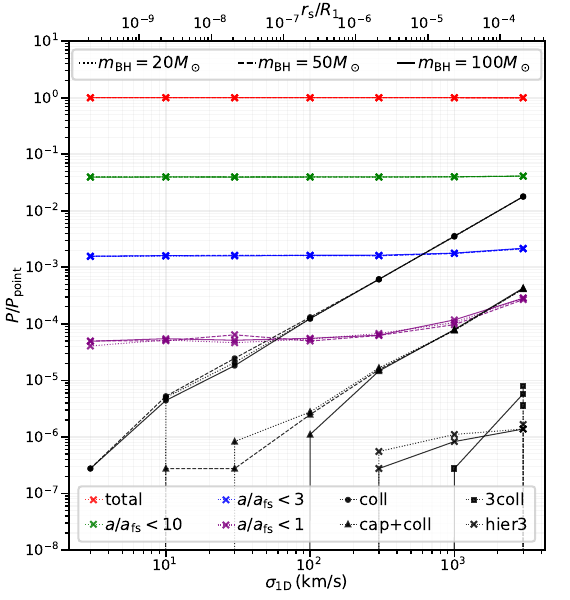}
\includegraphics[width=3in]{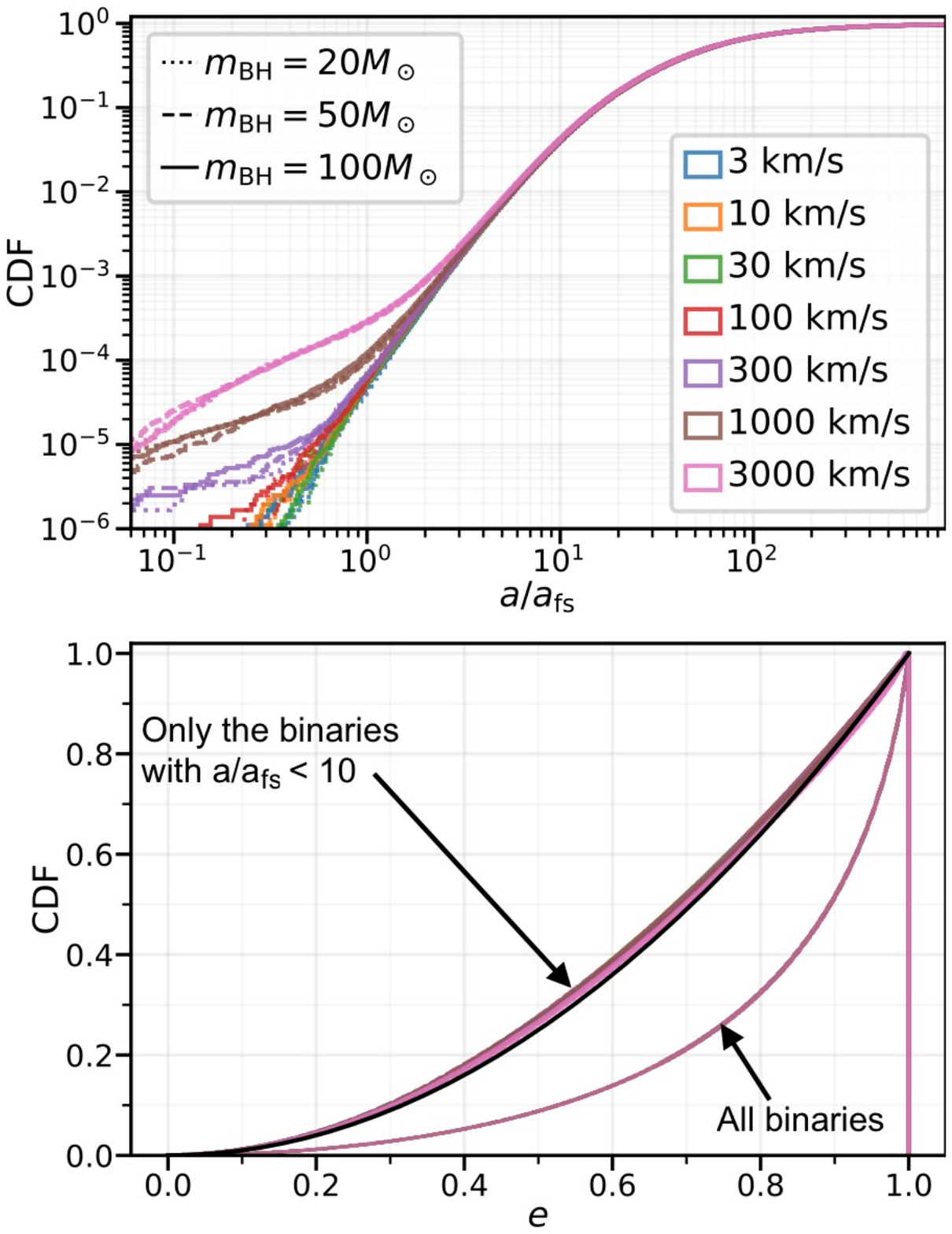}
\caption{3BBF distributions from scattering three equal-mass BHs, accounting for relativistic effects (post-Newtonian terms up to PN3.5) and varying the one-dimensional velocity dispersion $\sigma$.\\
\textit{Left}: 3BBF probability normalized by the total 3BBF probability (at ${\chi_1=10}$) in the point mass scenario (red curve in Figure~\ref{fig:Pijvschi1_AHcompare}). Line style denotes the mass of the BH, color represents hardness, and collision probabilities are in black. The black labels are: \textbf{coll}---one collision, \textbf{cap+coll}---collision and capture of the third body, \textbf{3coll}---all three bodies collide during scattering, \textbf{hier3}---formation of a hierarchical triple which lives for more than $10$ orbital periods of the outer tertiary before the inner binary merges. BH mergers are far less common than MS mergers under identical velocity profiles. Relativistic effects never impede 3BBF probabilities, instead augmenting hard 3BBF when $\sigma\gtrsim100\,\rm{km\,s^{-1}}$. \\
\textit{Right}: Cumulative SMA and eccentricity distributions for each choice of velocity dispersion. As $\sigma\rightarrow c$, GW emission encourages BH binaries to form with smaller SMA. Soft binaries (set of overlapping curves in the lower half of the eccentricity panel) are universally described by the same super-thermal eccentricity distribution as in the point mass case. Binaries with $a/a_{\rm fs}<10$ all closely follow a thermal eccentricity distribution (black curve), independent of the relativistic effects present during 3BBF.}
\label{fig:sigvseta_PN}
\end{center}
\end{figure*}

\subsection{Main-sequence Stars}\label{sec:main-sequence stars}
In Figure~\ref{fig:sigvseta_finitesize}, we present our findings for 3UB interactions between equal-mass MS stars for three different cases: ${m_{\rm i}/M_\odot = \{0.5,1,2\}}$. The left panel shows the 3BBF probabilities as a function of $\sigma$, colors again indicating binary hardness, with some additional black curves for collision probabilities (see caption). To better show the impact of finite size, we normalize the 3BBF probabilities by the total $\chi_1=10$ 3BBF probability in our equal point mass experiments from Section~\ref{sec:equalpointmasses} (the red curve in Figure~\ref{fig:Pijvschi1_AHcompare}). The right panels of Figure~\ref{fig:sigvseta_finitesize} show the corresponding SMA and eccentricity distributions. In the lower panel, we show both the eccentricity distributions for all binaries regardless of SMA (lower set of overlapping curves), as well as only the subset of binaries with $a<10 a_{\rm fs}$ (upper set of overlapping curves, predominantly found near the black curve representing a thermal distribution). Since it is dominated by wide binaries, the total binary population features  the same super-thermal eccentricity distribution found in our the equal point mass experiments. This distribution does not depend on $\sigma$, as evidenced by the extreme degree of overlap of these curves. Meanwhile, the subset of binaries with $a<10 a_{\rm fs}$ feature a nearly thermal eccentricity distribution, except at very high $\sigma\gtrsim 300\,{\rm km\, s}^{-1}$, where eccentricity instead skews significantly sub-thermal; solely due to eccentric binary periapse distances becoming comparable to MS stellar radii.

From the collision rates displayed in the left panel, it is immediately apparent that collisions are often orders of magnitude more common when compared to hard 3BBF $(P_{\rm coll}/P_{\rm hard} \in [10^0, 10^3])$. Yet, for all but the highest choices of $\sigma$ (lowest $R_1$), the inclusion of finite-size collisional effects does not appreciably alter the 3BBF probability from the simpler point mass scenario. Changes to hard 3BBF specifically (purple) are statistically insignificant until ${\sigma\gtrsim 30 \, \rm{km \, s^{-1}}}$, corresponding to $r_{\rm i }/R_1\gtrsim10^{-4}$. Notably, this already exceeds $\sigma$ at the center of typical globular clusters. Collisional suppression of 3BBF becomes relevant as the radius of the interaction volume, $R_1$, approaches $10^3\,R_\odot$. Changes to the SMA and eccentricity distributions at low $\sigma$ are similarly negligible when compared to the point mass limit. At speeds high enough that the point mass and finite size scenarios strongly deviate from each other, the SMA and eccentricity deviations increase rapidly with $\sigma$ (decreasing $R_1$). It is only when $\sigma \gtrsim 100\,{\rm km\,s}^{-1}$ that collisions severely hinder formation of especially hard and/or eccentric binaries. This may be relevant to nuclear star clusters, but not open or globular clusters. 

Fixing velocity and $\chi_1$, the radius $R_1$ of the interaction volume increases faster with mass than the radius $r_{\rm i}$ of MS stars. This can be seen from the proportion ${R_1/r_{\rm i} \propto m_{\rm i}^{2/5}}$ in Equations~(\ref{eq:impact_parameter}) and (\ref{eq:msradius}). Therefore, given identical velocity profiles, 3BBF involving more massive MS stars is less impeded by collisions---i.e., collisions in 3UB encounters are more frequent for lower-mass MS stars. The cross sections for two-body collisions and capture+collision events (two of the bodies colliding wth the collision product forming a binary with the third star) scale as ${P\propto\sigma^2\propto b_{90}^{-1}}$. The probability of all three bodies colliding scales as ${P\propto\sigma^4\propto b_{90}^{-2}}$. 

That collisions arising from 3UB encounters do not significantly reduce the 3BBF rate from the point mass case reveals a fundamental aspect of 3BBF physics. Namely, the overwhelming majority of initial configurations that align two-body trajectories into extremely tight, hyperbolic periapse passages do \textit{not} produce 3BBF in the point mass regime. Naturally, for finite-size bodies, such close passages result in collisions instead. This finding contradicts the intuition that if a tertiary body approaches two already strongly interacting bodies, a tight binary will form. We have no evidence that this occurs in the 3BBF animations we have generated of hard or soft binary formation. Instead, our results strongly suggest that all three bodies ``democratically'' participate in a perturbative binary formation process.

\subsection{Black Holes}\label{sec:black holes}
Figure~\ref{fig:sigvseta_PN} presents the outcomes of 3UB interactions between three equal-mass BHs, accounting for relativistic effects, in three separate cases: ${m_{\rm i}/M_\odot = \{20,50,100\}}$. Layout and line coloration/style are identical to Figure~\ref{fig:sigvseta_finitesize} for MS stars. All 3BBF probabilities in the left panel are again normalized by the total 3BBF probability for equal point masses (at ${\chi_1=10}$) in Section~\ref{sec:equalpointmasses}. As expected, accounting for relativistic effects negligibly changes the 3BBF probability from the Newtonian point mass scenario except in the case of hard 3BBF (purple) at high $\sigma \gtrsim 300\,{\rm km\,s^{-1}}$ ($r_s/R_1 \gtrsim 10^{-6}$). It is also clear from the collision rates (black) that BH mergers are far less likely than collisions between typical MS stars at identical velocities. This is a natural result of the minuscule physical cross-section for two-body GW capture compared to the (comparatively enormous) physical radius of MS stars. Using the GW capture radius, $r_{\rm p,GW}$, from \citet{Quinlan_1989}, we find that ${r_{\rm p,GW}/R_1\propto (\sigma/v)^{10/7}}$. This is precisely the scaling with $\sigma$ that we find for the probabilities of collision (\textbf{coll}; black circles) and capture+collision (\textbf{cap+coll}; black triangles).

Additionally, the scalings with $\sigma$ of the 3BBF and collision probabilities are independent of mass since $r_{\rm s}$ and $R_1$ both scale ${\propto m}$ with $r_{\rm i}/R_1 = 0.3 (\sigma/c)^2$. In other words, relativistic deviations from the point mass scenario scale only in powers of $v/c$. These relativistic deviations do not begin to peak above numerical noise until $\sigma\gtrsim 100 \,\rm{km \, s^{-1}}$ (i.e., $\sigma / c \gtrsim 10^{-4}$), observable in the subtle boost to the hardest portion of the SMA distribution, becoming more dramatic as $\sigma/c \rightarrow 1$. We find no deviation in eccentricity distributions in comparison to point mass interactions. Thus, hard 3BBF with BHs is well-described by a thermal eccentricity distribution. 

We are also excited to report a small probability in which three unbound BHs may undergo a 3UB interaction that produces a hierarchical triple via GW emission. Though most capture+collision end-states (\textbf{cap+coll}) involve the formation of a short-lived triple BH system, we only classify outcomes as a hierarchical triple (\textbf{hier3}; black crosses) when the hierarchy survives for $\tau{\gtrsim}10$ orbital periods of the outer tertiary before the inner binary merges. A binary BH containing a second-generation BH merger product is always left behind in these equal-mass, zero-spin scenarios due to the nonexistence of GW recoil kicks in such a case. However, we stress that this is an extremely rare occurrence and is poorly resolved, even for environments with a local velocity dispersion in excess of $1000\, \rm{km \, s^{-1}}$.

\begin{figure}
\includegraphics[width=3.35in]{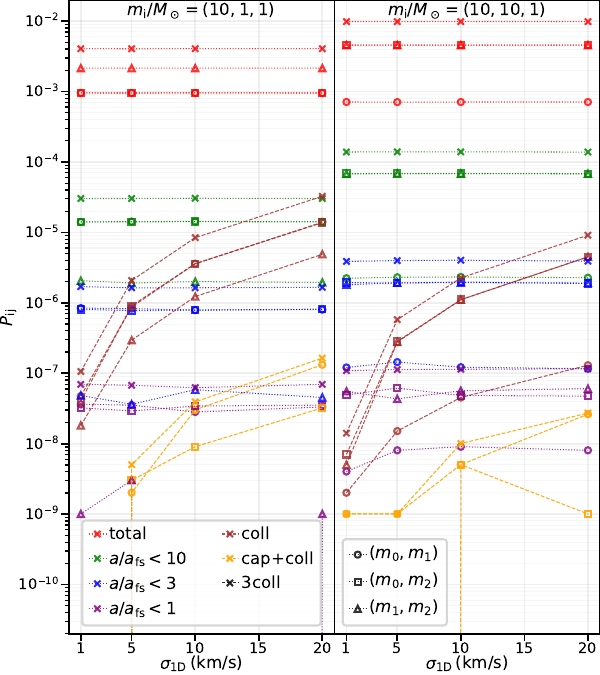}
\caption{3BBF probabilities (non-normalized) between (\textit{left}) a BH and two MS stars and (\textit{right}) two BHs and an MS star with $\chi_1=10$. All BHs (initially non-spinning) have mass ${10\, M_\odot}$ and all MS stars have mass ${1\, M_\odot}$. These scattering experiments are thus the finite-sized, relativistic equivalent of the earlier simulation sets for unequal masses with ${q_{\rm i} = \{1,0.1,0.1\}}$ (left) and ${q_{\rm i} = \{1,1,0.1\}}$ (right), respectively. Specific binary pairing combinations are identified by shape as in Figure~\ref{fig:sigvseta_q}. Colors distinguish 3BBF resulting in different binary hardness, as well as several different collision probabilities (labeled as in Figures~\ref{fig:sigvseta_finitesize} and \ref{fig:sigvseta_PN}).  We find that 3BBF rates are largely unchanged compared to the Newtonian point mass limit, but direct collisions occur more frequently as the local velocity dispersion increases. Tidal disruption events should be even more frequent since they have a larger cross section compared to direct BH--star collisions.
}
\label{fig:sigvseta_massvelPN}
\end{figure}

\begin{figure}
\includegraphics[width=3.35in]{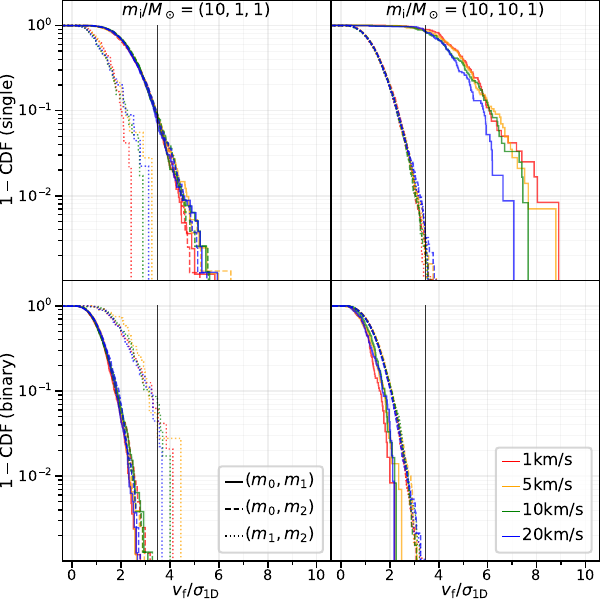}
\caption{Survival function $(1 - C(v_{\rm f}/\sigma))$ for the final velocity $(v_{\rm f}/\sigma)$ of the catalyzing single (\textit{top row}) and binary (\textit{bottom row}). These distributions include only the binaries formed with $a_{\rm ij}/a_{\rm fs, ij}<3$. Color and line style illustrate the velocity dispersion and specific mass pairing, respectively. The vertical black line at $v_{\rm f}/\sigma= 2\sqrt{3}$ denotes the escape velocity from the core of a Plummer cluster.}
\label{fig:vk_msbh}
\end{figure}

\subsection{Main-sequence Stars and Black Holes}\label{sec:msandbh}
A key application of 3UB encounters is the scenario involving MS stars $({\sim}1\, M_\odot)$ and stellar-mass BHs $({\sim}10 \,M_\odot)$ within the cores of dense star clusters. \citet{Weatherford_2023a} recently demonstrated that such encounters dominate 3BBF at most times in typical Milky Way globular clusters (those retaining significant BH populations). They also found that such interactions may dominate high-speed ejection $(\Delta v_{\rm k}> \sigma)$ from such clusters due to the recoil experienced by the leftover single, expected to typically be a low-mass MS star. However, the frequency of high-speed ejection and the identity of the ejected body (i.e., high-mass or low-mass object, single or binary) may be conditional on previously unexplored unequal-mass, finite-size, and relativistic effects.

In Figure~\ref{fig:sigvseta_massvelPN}, we reaffirm that physical collisions, despite being more frequent than hard 3BBF, do not significantly suppress hard 3BBF when compared to the point mass scenario for the unequal masses (and velocity dispersions; $\sigma \in [1,20]\,\rm{km\,s}^{-1}$) explored here. This result is consistent with our findings in Sections~\ref{sec:main-sequence stars}~and~\ref{sec:black holes} that high-$\sigma$ environments are necessary to significantly suppress hard 3BBF among equal-mass MS stars (${\sigma\gtrsim 30\,\rm{km\,s^{-1}}}$) or BHs (${\sigma\gtrsim 100\,\rm{km\,s^{-1}}}$).

BH--star 3BBF rates for both mass combinations are unchanged compared to the Newtonian point mass scenario (see Figure~\ref{fig:sigvseta_q}), but the collision rate is substantial. As we demonstrated in Section~\ref{sec:main-sequence stars}, initial configurations that produce collisions are more frequent than hard 3BBF when scattering finite-sized bodies. However, these initial states are non-degenerate for typical cluster velocity dispersions; i.e., an initial state that produces a hard binary in the point mass regime does not necessarily produce a collision during an identical encounter with finite-size bodies. An unexpected consequence follows if we consider that the tidal disruption radius, ${r_{\rm t}=r_* \left(m_{\rm BH}/m_{*} \right)^{1/3}}$, is about $2.2$ times larger than the BH--star collision radius between a $10\, M_\odot$ BH and a ${1 \, M_\odot}$ MS star. Namely, the number of BH--star tidal disruption events may be between {$2$ to $5$} times higher than the direct collision rate, contingent on the subtleties of 3UB gravitational focusing. While direct BH--star collisions do not meaningfully suppress the hard 3BBF rate, we are uncertain whether tidal disruption events would yield significant suppression. Regardless, these results suggest 3UB encounters may provide a substantial boost to the predicted frequency of transient tidal/collisional phenomena in the cores of dense star clusters.

In Figure~\ref{fig:vk_msbh}, we also examine the final velocity $v_{\rm{f}}$ of binaries and the leftover (catalyzing) singles produced by 3BBF from the above setup (i.e., BHs of mass $10\,M_\odot$ and MS stars of mass $1\,M_\odot$). Note the velocities shown are in the frame of the interaction volume, akin to the global reference frame of the host environment, not the center-of-mass frame of the three bodies. The primary advantage is the direct translation to the final velocity of each body within its host environment. We also only include in the analysis binaries formed with ${a_{\rm ij}/a_{\rm fs} \leq 3}$---i.e., hard and slightly softer binaries. 

Given our assumption that the bodies' velocities are drawn from the same distribution regardless of mass, we find it is excessively rare that newly formed BH--MS or BH--BH \textit{binaries} have a final velocity in excess of the escape velocity ($v_{\rm esc}=2 \sqrt{3} \sigma$) of a Plummer core with the same characteristic velocity dispersion. \dva{Even in the favorable case where a (heavy) BH catalyzes the formation of a (light) MS--MS binary with $a_{\rm 12}/a_{\rm fs} \leq 3$, ejection of the binary only occurs ${<}5\%$ of the time. Notably, Figure~\ref{fig:sigvseta_massvelPN} shows that the MS--MS pairing in such a case is also $10$ times \textit{less} likely than a BH--MS pairing, for which ejection of the binary would be even rarer.} \citet{Weatherford_2023a} similarly predicted that the energy release from 3BBF alone is almost never directly responsible for the ejection of a binary. Notably, the simplified 3BBF prescription from the \texttt{CMC} code used in that study automatically pairs the most massive bodies, artificially making binary ejection more unlikely than the otherwise small fraction found here.

From simple momentum conservation, the prospect of low-mass catalysts experiencing high-speed kicks from 3BBF is much more promising. \dva{When a MS star catalyzes the formation of a BH--MS or BH--BH binary with $a_{\rm ij}\leq3 a_{\rm fs,ij}$, ${\sim}10\%$ and ${\sim}90\%$} of MS star catalysts have final velocities $v_{\rm f}>v_{\rm esc}$, respectively. \dva{Hard 3BBF can lead to even higher ejection speeds for single stars as $v_{\rm f}\propto \sigma$ if we only consider point-mass Newtonian mechanics. As discussed in Section~\ref{sec:finitesize_PN}, by additionally considering finite-size effects, direct collisions between stars with masses of ${\sim}1\,M_\odot$ begin to reduce 3BBF for velocity dispersions between $30$--$50 \, \rm{km \, s^{-1}}$ and eliminate hard 3BBF at $\sigma > 100 \, \rm{km \, s^{-1}}$. This reduction in 3BBF rates due to collisions is less pronounced for more massive bodies since the size of the interaction volume increases faster than the physical radii of interacting stars.}

At an order-of-magnitude level, these results are consistent with the velocity distribution from MS star ejecta via 3BBF in \texttt{CMC} \citep[][]{Weatherford_2023a}. However, there are a variety of complexities that need to be further explored to make a more rigorous comparison. For example, the initial velocities of the bodies in Figure~\ref{fig:vk_msbh} are sampled from the local velocity dispersion (typical velocity), while successful 3BBF ejecta from \citet{Weatherford_2023a} should naturally be skewed to (higher) initial velocities already near to $v_{\rm esc}$. Furthermore, Figure~\ref{fig:vk_msbh} does not account for species of different mass having different $\sigma$; nor does it account for the full distribution of mass ratios and relative velocity ratios found in typical 3UB encounters in the core of a dense star cluster. We leave a more rigorous analysis of such considerations to future work.

Finally, as mentioned earlier, tidal physics will likely modify the final velocity of the catalyst. For example, \citet{Kremer_2022} demonstrated that velocity kicks applied to stellar remnants following tidal disruption events (due to asymmetric mass loss) may exceed ${\sim}200\,\rm{km\, s^{-1}}$ when the mass of the BH is at least $10$ times greater than the star's mass. So the prospect of impulsive acceleration to high speed during a 3UB encounter is likely much higher than suggested by our results, which account only for (post-)Newtonian corrections. Along with Figure~\ref{fig:vk_msbh}, we find that 3UB encounters should contribute to the ejection of runaway stars (and possibly even hypervelocity stars) from globular clusters, supporting \citet{Weatherford_2023a}. Conclusively demonstrating that 3BBF dominates the production of such ejecta over other binary-mediated channels requires further numerical examination of the 3BBF rate for realistic velocity and mass distributions in globular clusters. This is especially true when considering hypervelocity ejecta, which fundamentally would result from encounters deep in the tail of the 3BBF kick velocity distribution.

\vspace{0.5cm}
\section{Summary and Future Work}\label{sec:conclusion}
\subsection{Summary}\label{sec:summary}
We have explored the formation of new binaries from three unbound bodies in greater detail than any prior efforts, including the first study of 3BBF to feature unequal masses, finite-size effects, and post-Newtonian effects. Specifically, we study 3BBF by running ${>}3\times10^{10}$ three-body scattering experiments with the state-of-the-art direct $N$-body integrator \textsc{Tsunami}. After reproducing the canonical 3BBF scattering experiments of \AH{}, we correct an oversight in their algorithm \citep[adopted from][]{Agekyan_1971} related to spherical symmetry. With this correction and a new analytic encounter rate for 3BBF, we compute a hard 3BBF rate that agrees well with \citet{Goodman_1993} in the case of equal-mass bodies; see Equation~(\ref{eq:estimated_formation_rate}). We also confirm that 3BBF is almost exclusively an impulsive phenomenon. Resonant encounters leading to 3BBF are extremely disfavored due to the total positive energy of three unbound bodies, preventing a three-body bound state from occurring without dissipation (e.g., gravitational radiation or collisions). Additional key results are as follows:

\begin{enumerate}
\item Newly formed binaries from 3BBF are overwhelming soft, in agreement with \AH{} and \citet{Goodman_1993}. We find that the cumulative distribution of binary semi-major axis, $a$, scales as $a^3$---notably shallower than the $a^{5.5}$ scaling of 3BBF recipes in Monte Carlo star cluster modeling \citep[e.g., \texttt{CMC};][]{CMCRelease}. Typical SMA also grows with dimensionless impact parameter $\chi_1$, so larger interaction volumes lead to both wider binaries and higher 3BBF rates (since 3UB encounter rates scale as $\chi_1^5$). This has dramatic implications for star cluster dynamics since soft binaries are often assumed to disrupt too quickly to affect cluster evolution. Yet \citet{Goodman_1993} estimate ${\gtrsim}90\%$ of long-lived binaries form soft; they are simply the rare few of many soft binaries from 3BBF that survived and hardened. So neglect of soft 3BBF in cluster modeling may warrant re-examination.
\item Wide binaries from 3BBF have super-thermal eccentricities. Since most binaries form soft, this implies that soft 3BBF may produce the eccentric wide binaries observed with \textit{Gaia} \citep[e.g.,][]{Tokovinin_2020,Hwang_2022}.
Star clusters may therefore be robust sources of eccentric wide binaries. In particular, such binaries may form via 3BBF interactions in short-lived or dissolving clusters \citep[though this likely results in a thermal eccentricity distribution, e.g.,][]{Kouwenhoven2010}, or perhaps in the extended tidal tails of more massive clusters. Although the local density in tidal tails is significantly lower than in the core of a star cluster, the highly correlated epicyclic trajectories of slow escapers in tidal tails may produce exceptionally low relative velocities between neighboring bodies conducive to wide 3BBF; recall the 3BBF rate depends much more steeply on relative velocity than on density. Survival of newly formed wide binaries and their deposition into the Galactic field may also be easier in this case, and the extreme velocity anistropy in tidal tails may result in significantly different semi-major axis or eccentricity distributions than expected from an isotropic assumption for 3BBF (as in this paper). We shall examine such prospects in future work.

\item Independent of mass ratio, the eccentricity distribution of hard binaries formed through 3BBF is universally well-described by the classic thermal distribution \citep[e.g.,][]{Jeans_1919,Heggie_1975}. This likely occurs because an isotropically distributed sea of gravitating bodies will fully explore phase space, analogous to how resonant binary--single encounters fully explore phase space  in the aforementioned texts. The $a^3$ scaling of the cumulative distribution for SMA also holds for hard binaries in most cases; a key exception is the (unlikely) pairing of two massive bodies by a low-mass catalyst.
\item Exploring unequal-mass 3BBF for the first time, our results refute the common assumption that the two most massive bodies are the most likely to pair \citep[e.g.,][]{Morscher_2013}. Instead, the two least massive bodies pair most frequently in soft/wide binaries while the most massive plus least massive bodies pair most frequently in hard binaries. Only for mass ratios near unity are the two most massive bodies likely to pair (up to ${\approx}40\%$ of hard 3BBFs).
\item Physical stellar collisions are a more likely outcome of a 3UB encounter than hard binary formation for MS stars with velocities ${\sigma>1\,\rm{km\,s}^{-1}}$. Yet the collisions do \text{not} significantly suppress 3BBF among MS stars (relative to the point mass limit) at velocity dispersions $\lesssim 30\,\rm{km\,s}^{-1}$. This implies that the initial states leading to stellar collisions in 3UB interactions are largely separate from those leading to hard binary formation.
\item Post-Newtonian effects can promote hard binary BH formation through gravitational wave emission during close high-speed approach. Yet such enhancement (and prevalence of BH mergers) is only significant for $\sigma \gtrsim 100\,\rm{km\,s}^{-1}$, likely only relevant in nuclear star clusters. In such high-$\sigma$ environments, the probability of forming short-lived hierarchical triples through 3UB BH scattering can also be significant, surpassing the hard 3BBF probability. Formation of longer-lived triples (surviving for at least several orbits of the outer tertiary) remains rarer than hard 3BBF, even when $\sigma \gtrsim 1000\,\rm{km\,s}^{-1}$.
\item The above results hold for mixed-species interactions between stellar-mass BHs and MS stars, the dominant type of 3UB encounters in star cluster models \citep{Weatherford_2023a}. For typical masses and velocities in globular clusters, neither direct collisions nor post-Newtonian effects significantly alter 3BBF rates in this case relative to the limit of Newtonian point masses. As with MS stars alone, BH--MS collisions in these mixed-species encounters are much more common than hard 3BBF, but do not significantly suppress 3BBF. Since the cross section for a tidal disruption event (TDE) is larger than for a direct BH--MS collision, 3UB encounters in star clusters may be a significant source of TDEs.
\item We confirm that 3BBF can eject bodies from star clusters at speeds at least a few times their central escape velocity (depending on the mass ratios in the encounter). In particular, when a MS star catalyzes the production of a typical hard BH--MS or BH--BH binary, ${\sim}10\%$ and ${\sim}90\%$, respectively, of the MS catalysts exit the 3UB encounter fast enough to escape their host cluster. 3BBF in star clusters may therefore contribute to runaway stars in the Galactic halo, but at rates that remain uncertain pending future work that more closely examines 3BBF in a background environment with fully realistic mass and velocity distributions. Tidal disruption physics may further enhance high-speed ejection from 3BBF since asymmetric mass loss imparts an additional kick to the stripped star \citep{Kremer_2022}. Newly formed \textit{binaries} rarely exit 3BBF events with sufficient speed to escape from the center of a star cluster, even when the catalyst is much more massive than the binary.

\item \dva{Finally, our results agree with an independent analytical investigation of 3BBF by \citet{Ginat_2024_3BBF}, submitted during the review process of this work. In particular, \citet{Ginat_2024_3BBF} developed a statistical analytic theory investigating the statistics and distributions of orbital parameters of equal-mass 3BBF and agree that 3BBF is a promising source of soft, wide eccentric binaries featuring super-thermal eccentricity distributions. They also corroborated that the 3BBF probability for binaries formed with SMA less than their strong interaction region (i.e., $a < R_1$) scales as $P_{\rm ij}\propto \chi_1^{-4.5}$ and that these binaries have an approximately thermal eccentricity distribution.}

\end{enumerate}

\subsection{Future Work}\label{sec:future_work}

Despite touching on some of the more extreme regimes of the 3UB parameter space in this work, the overwhelming majority are yet to be explored (e.g., varying velocity ratios, non-isotropic environments, energy partitioning). Given the number of other considerations investigated here, these regimes are beyond the scope of this work, but may dramatically impact binary properties. 

Proper treatment of tidal physics in the close passages of stars, including tidal disruption events, in 3UB scattering may be impactful and requires focused study. In particular, our results in Section~\ref{sec:msandbh} demonstrate that direct BH--star collisions do not meaningfully reduce the rate of hard binary formation, yet they are a \textit{more common} outcome. The complex interplay between tidal physics and hard binary formation is yet to be explored, but we do know the tidal disruption rate will be at least twice as frequent as the collision rate---if only due to gravitational focusing. It immediately follows that tidal disruption events are likely highly prevalent in 3UB encounters within dense stellar clusters. Additionally, future studies will be targeted towards specific environments, enabling us to make concrete and practical environment-specific 3BBF rate predictions.

In total, our results should serve as a reminder that the 3UB problem is largely unexplored. Expectations concerning populations of dynamically assembled binaries may change significantly when a proper treatment of arbitrary mass-ratio 3UB scattering is incorporated into Monte Carlo star cluster modeling codes such as \texttt{CMC}. While 3BBF features obvious applications in constraining the history of dynamically assembled compact-object binaries and their subsequent mergers observable through gravitational waves, the potential for enhancing our understanding of stellar binary formation provides further incentive for renewing investigations into 3BBF as a critical topic in dynamical astrophysics.

\section*{Acknowledgements}
This work was supported by NSF Grant AST-2108624 at Northwestern University. A.A.T. acknowledges support from the European Union’s Horizon 2020 and Horizon Europe research and innovation programs under the Marie Sk\l{}odowska-Curie grant agreements No.~847523 and 101103134. We thank Barry Ginat for informative and positive conversations regarding their analytic analysis of 3BBF and Yoram Lithwick, Mike Zevin, Kyle Kremer, Chris Hamilton, and Jeff Andrews for insightful discussions regarding numerical sampling methods and astrophysical implications. This work was supported through the computational resources and staff contributions provided for the Quest high-performance computing facility at Northwestern University. Quest is jointly supported by the Office of the Provost, the Office for Research, and Northwestern University Information Technology. DA also acknowledges support from a CIERA Board of Visitors Fellowship and the computing resources at CIERA funded by NSF Grant~PHY-1726951.

\bibliography{3BBF}{}

\begin{thebibliography}{}
\expandafter\ifx\csname natexlab\endcsname\relax\def\natexlab#1{#1}\fi
\providecommand{\url}[1]{\href{#1}{#1}}
\providecommand{\dodoi}[1]{doi:~\href{http://doi.org/#1}{\nolinkurl{#1}}}
\providecommand{\doeprint}[1]{\href{http://ascl.net/#1}{\nolinkurl{http://ascl.net/#1}}}
\providecommand{\doarXiv}[1]{\href{https://arxiv.org/abs/#1}{\nolinkurl{https://arxiv.org/abs/#1}}}

\bibitem[{{Aarseth} \& {Heggie}(1976)}]{Aarseth_1976}
{Aarseth}, S.~J., \& {Heggie}, D.~C. 1976, \aap, 53, 259

\bibitem[{{Agekyan} \& {Anosova}(1971)}]{Agekyan_1971}
{Agekyan}, T.~A., \& {Anosova}, Z.~P. 1971, \sovast, 15, 411

\bibitem[{{Arca Sedda} {et~al.}(2023){Arca Sedda}, {Kamlah}, {Spurzem}, {Rizzuto}, {Naab}, {Giersz}, \& {Berczik}}]{ArcaSedda2023}
{Arca Sedda}, M., {Kamlah}, A. W.~H., {Spurzem}, R., {et~al.} 2023, \mnras, 526, 429, \dodoi{10.1093/mnras/stad2292}

\bibitem[{{Atallah} {et~al.}(2023){Atallah}, {Trani}, {Kremer}, {Weatherford}, {Fragione}, {Spera}, \& {Rasio}}]{Atallah_2023}
{Atallah}, D., {Trani}, A.~A., {Kremer}, K., {et~al.} 2023, \mnras, 523, 4227, \dodoi{10.1093/mnras/stad1634}

\bibitem[{{Bacon} {et~al.}(1996){Bacon}, {Sigurdsson}, \& {Davies}}]{Bacon1996}
{Bacon}, D., {Sigurdsson}, S., \& {Davies}, M.~B. 1996, \mnras, 281, 830, \dodoi{10.1093/mnras/281.3.830}

\bibitem[{{Banerjee} {et~al.}(2010){Banerjee}, {Baumgardt}, \& {Kroupa}}]{Banerjee2010}
{Banerjee}, S., {Baumgardt}, H., \& {Kroupa}, P. 2010, \mnras, 402, 371, \dodoi{10.1111/j.1365-2966.2009.15880.x}

\bibitem[{{Binney} \& {Tremaine}(2008)}]{Binney_2008}
{Binney}, J., \& {Tremaine}, S. 2008, {Galactic Dynamics: Second Edition}

\bibitem[{{Chernoff} \& {Huang}(1996)}]{ChernoffHuang1996}
{Chernoff}, D.~F., \& {Huang}, X. 1996, in Dynamical Evolution of Star Clusters: Confrontation of Theory and Observations, ed. P.~{Hut} \& J.~{Makino}, Vol. 174, 263

\bibitem[{{Demircan} \& {Kahraman}(1991)}]{Demircan_1991}
{Demircan}, O., \& {Kahraman}, G. 1991, \apss, 181, 313, \dodoi{10.1007/BF00639097}

\bibitem[{{Fabian} {et~al.}(1975){Fabian}, {Pringle}, \& {Rees}}]{Fabian_1975}
{Fabian}, A.~C., {Pringle}, J.~E., \& {Rees}, M.~J. 1975, \mnras, 172, 15, \dodoi{10.1093/mnras/172.1.15P}

\bibitem[{{Fregeau} {et~al.}(2004{\natexlab{a}}){Fregeau}, {Cheung}, {Portegies Zwart}, \& {Rasio}}]{Fregeau2004}
{Fregeau}, J.~M., {Cheung}, P., {Portegies Zwart}, S.~F., \& {Rasio}, F.~A. 2004{\natexlab{a}}, \mnras, 352, 1, \dodoi{10.1111/j.1365-2966.2004.07914.x}

\bibitem[{{Fregeau} {et~al.}(2004{\natexlab{b}}){Fregeau}, {Cheung}, {Portegies Zwart}, \& {Rasio}}]{2004MNRAS.352....1F}
---. 2004{\natexlab{b}}, \mnras, 352, 1, \dodoi{10.1111/j.1365-2966.2004.07914.x}

\bibitem[{{Fregeau} {et~al.}(2003){Fregeau}, {G{\"u}rkan}, {Joshi}, \& {Rasio}}]{Fregeau2003}
{Fregeau}, J.~M., {G{\"u}rkan}, M.~A., {Joshi}, K.~J., \& {Rasio}, F.~A. 2003, \apj, 593, 772, \dodoi{10.1086/376593}

\bibitem[{{Freitag} \& {Benz}(2001)}]{FreitagBenz2001}
{Freitag}, M., \& {Benz}, W. 2001, \aap, 375, 711, \dodoi{10.1051/0004-6361:20010706}

\bibitem[{{Generozov} {et~al.}(2018){Generozov}, {Stone}, {Metzger}, \& {Ostriker}}]{Generozov_2018}
{Generozov}, A., {Stone}, N.~C., {Metzger}, B.~D., \& {Ostriker}, J.~P. 2018, \mnras, 478, 4030, \dodoi{10.1093/mnras/sty1262}

\bibitem[{{Ginat} \& {Perets}(2024)}]{Ginat_2024_3BBF}
{Ginat}, Y.~B., \& {Perets}, H.~B. 2024, arXiv e-prints, arXiv:2404.08040, \dodoi{10.48550/arXiv.2404.08040}

\bibitem[{{Goodman} \& {Hernquist}(1991)}]{GoodmanHernquist1991}
{Goodman}, J., \& {Hernquist}, L. 1991, \apj, 378, 637, \dodoi{10.1086/170464}

\bibitem[{{Goodman} \& {Hut}(1993)}]{Goodman_1993}
{Goodman}, J., \& {Hut}, P. 1993, \apj, 403, 271, \dodoi{10.1086/172200}

\bibitem[{{Hamilton} \& {Modak}(2023)}]{Hamilton_2023}
{Hamilton}, C., \& {Modak}, S. 2023, arXiv e-prints, arXiv:2311.04352, \dodoi{10.48550/arXiv.2311.04352}

\bibitem[{{Healy} \& {Lousto}(2018)}]{Healy_2018}
{Healy}, J., \& {Lousto}, C.~O. 2018, \prd, 97, 084002, \dodoi{10.1103/PhysRevD.97.084002}

\bibitem[{{Heggie} \& {Hut}(2003)}]{Heggie_2003}
{Heggie}, D., \& {Hut}, P. 2003, {The Gravitational Million-Body Problem: A Multidisciplinary Approach to Star Cluster Dynamics}

\bibitem[{{Heggie}(1975)}]{Heggie_1975}
{Heggie}, D.~C. 1975, \mnras, 173, 729, \dodoi{10.1093/mnras/173.3.729}

\bibitem[{{Hills}(1990)}]{Hills_1990}
{Hills}, J.~G. 1990, \aj, 99, 979, \dodoi{10.1086/115388}

\bibitem[{{Hut}(1985)}]{Hut1985}
{Hut}, P. 1985, in Dynamics of Star Clusters, ed. J.~{Goodman} \& P.~{Hut}, Vol. 113, 231--247

\bibitem[{{Hut} \& {Inagaki}(1985)}]{HutInagaki1985}
{Hut}, P., \& {Inagaki}, S. 1985, \apj, 298, 502, \dodoi{10.1086/163636}

\bibitem[{{Hut} {et~al.}(1992){Hut}, {McMillan}, \& {Romani}}]{Hut1992}
{Hut}, P., {McMillan}, S., \& {Romani}, R.~W. 1992, \apj, 389, 527, \dodoi{10.1086/171229}

\bibitem[{{Hwang} {et~al.}(2022){Hwang}, {Ting}, \& {Zakamska}}]{Hwang_2022}
{Hwang}, H.-C., {Ting}, Y.-S., \& {Zakamska}, N.~L. 2022, \mnras, 512, 3383, \dodoi{10.1093/mnras/stac675}

\bibitem[{{Ivanova} {et~al.}(2005){Ivanova}, {Belczynski}, {Fregeau}, \& {Rasio}}]{Ivanova_2005}
{Ivanova}, N., {Belczynski}, K., {Fregeau}, J.~M., \& {Rasio}, F.~A. 2005, \mnras, 358, 572, \dodoi{10.1111/j.1365-2966.2005.08804.x}

\bibitem[{{Ivanova} {et~al.}(2010){Ivanova}, {Chaichenets}, {Fregeau}, {Heinke}, {Lombardi}, \& {Woods}}]{Ivanova_2010}
{Ivanova}, N., {Chaichenets}, S., {Fregeau}, J., {et~al.} 2010, \apj, 717, 948, \dodoi{10.1088/0004-637X/717/2/948}

\bibitem[{{Jeans}(1919)}]{Jeans_1919}
{Jeans}, J.~H. 1919, \mnras, 79, 408, \dodoi{10.1093/mnras/79.6.408}

\bibitem[{{Joshi} {et~al.}(2001){Joshi}, {Nave}, \& {Rasio}}]{Joshi2001}
{Joshi}, K.~J., {Nave}, C.~P., \& {Rasio}, F.~A. 2001, \apj, 550, 691, \dodoi{10.1086/319771}

\bibitem[{{Joshi} {et~al.}(2000){Joshi}, {Rasio}, \& {Portegies Zwart}}]{Joshi2000}
{Joshi}, K.~J., {Rasio}, F.~A., \& {Portegies Zwart}, S. 2000, \apj, 540, 969, \dodoi{10.1086/309350}

\bibitem[{{Kouwenhoven} {et~al.}(2010){Kouwenhoven}, {Goodwin}, {Parker}, {Davies}, {Malmberg}, \& {Kroupa}}]{Kouwenhoven2010}
{Kouwenhoven}, M.~B.~N., {Goodwin}, S.~P., {Parker}, R.~J., {et~al.} 2010, \mnras, 404, 1835, \dodoi{10.1111/j.1365-2966.2010.16399.x}

\bibitem[{{Kremer} {et~al.}(2022){Kremer}, {Lombardi}, {Lu}, {Piro}, \& {Rasio}}]{Kremer_2022}
{Kremer}, K., {Lombardi}, J.~C., {Lu}, W., {Piro}, A.~L., \& {Rasio}, F.~A. 2022, \apj, 933, 203, \dodoi{10.3847/1538-4357/ac714f}

\bibitem[{{Kremer} {et~al.}(2021{\natexlab{a}}){Kremer}, {Piro}, \& {Li}}]{Kremer_2021}
{Kremer}, K., {Piro}, A.~L., \& {Li}, D. 2021{\natexlab{a}}, \apjl, 917, L11, \dodoi{10.3847/2041-8213/ac13a0}

\bibitem[{{Kremer} {et~al.}(2021{\natexlab{b}}){Kremer}, {Rui}, {Weatherford}, {Chatterjee}, {Fragione}, {Rasio}, {Rodriguez}, \& {Ye}}]{Kremer_2021_WDs}
{Kremer}, K., {Rui}, N.~Z., {Weatherford}, N.~C., {et~al.} 2021{\natexlab{b}}, \apj, 917, 28, \dodoi{10.3847/1538-4357/ac06d4}

\bibitem[{{Kulkarni} {et~al.}(1993){Kulkarni}, {Hut}, \& {McMillan}}]{Kulkarni1993}
{Kulkarni}, S.~R., {Hut}, P., \& {McMillan}, S. 1993, \nat, 364, 421, \dodoi{10.1038/364421a0}

\bibitem[{{Lousto} \& {Zlochower}(2013)}]{Lousto_2013}
{Lousto}, C.~O., \& {Zlochower}, Y. 2013, \prd, 87, 084027, \dodoi{10.1103/PhysRevD.87.084027}

\bibitem[{{Maoz} {et~al.}(2014){Maoz}, {Mannucci}, \& {Nelemans}}]{Maoz_2014}
{Maoz}, D., {Mannucci}, F., \& {Nelemans}, G. 2014, \araa, 52, 107, \dodoi{10.1146/annurev-astro-082812-141031}

\bibitem[{{Mar{\'\i}n Pina} \& {Gieles}(2023)}]{Pina_2023}
{Mar{\'\i}n Pina}, D., \& {Gieles}, M. 2023, arXiv e-prints, arXiv:2308.10318, \dodoi{10.48550/arXiv.2308.10318}

\bibitem[{{McMillan}(1986)}]{McMillan1986}
{McMillan}, S.~L.~W. 1986, \apj, 306, 552, \dodoi{10.1086/164365}

\bibitem[{{Mikkola} \& {Aarseth}(1993)}]{kschain}
{Mikkola}, S., \& {Aarseth}, S.~J. 1993, Celestial Mechanics and Dynamical Astronomy, 57, 439, \dodoi{10.1007/BF00695714}

\bibitem[{{Morscher} {et~al.}(2015){Morscher}, {Pattabiraman}, {Rodriguez}, {Rasio}, \& {Umbreit}}]{Morscher_2015}
{Morscher}, M., {Pattabiraman}, B., {Rodriguez}, C., {Rasio}, F.~A., \& {Umbreit}, S. 2015, \apj, 800, 9, \dodoi{10.1088/0004-637X/800/1/9}

\bibitem[{{Morscher} {et~al.}(2013){Morscher}, {Umbreit}, {Farr}, \& {Rasio}}]{Morscher_2013}
{Morscher}, M., {Umbreit}, S., {Farr}, W.~M., \& {Rasio}, F.~A. 2013, \apjl, 763, L15, \dodoi{10.1088/2041-8205/763/1/L15}

\bibitem[{{O'Leary} {et~al.}(2006){O'Leary}, {Rasio}, {Fregeau}, {Ivanova}, \& {O'Shaughnessy}}]{OLeary2006}
{O'Leary}, R.~M., {Rasio}, F.~A., {Fregeau}, J.~M., {Ivanova}, N., \& {O'Shaughnessy}, R. 2006, \apj, 637, 937, \dodoi{10.1086/498446}

\bibitem[{{Pe{\~n}arrubia}(2021)}]{Pen_2021}
{Pe{\~n}arrubia}, J. 2021, \mnras, 501, 3670, \dodoi{10.1093/mnras/staa3700}

\bibitem[{{Quinlan} \& {Shapiro}(1989)}]{Quinlan_1989}
{Quinlan}, G.~D., \& {Shapiro}, S.~L. 1989, \apj, 343, 725, \dodoi{10.1086/167745}

\bibitem[{{Rodriguez} {et~al.}(2019){Rodriguez}, {Zevin}, {Amaro-Seoane}, {Chatterjee}, {Kremer}, {Rasio}, \& {Ye}}]{Rodriguez_2019}
{Rodriguez}, C.~L., {Zevin}, M., {Amaro-Seoane}, P., {et~al.} 2019, \prd, 100, 043027, \dodoi{10.1103/PhysRevD.100.043027}

\bibitem[{{Rodriguez} {et~al.}(2022){Rodriguez}, {Weatherford}, {Coughlin}, {Amaro-Seoane}, {Breivik}, {Chatterjee}, {Fragione}, {K{\i}ro{\u{g}}lu}, {Kremer}, {Rui}, {Ye}, {Zevin}, \& {Rasio}}]{CMCRelease}
{Rodriguez}, C.~L., {Weatherford}, N.~C., {Coughlin}, S.~C., {et~al.} 2022, \apjs, 258, 22, \dodoi{10.3847/1538-4365/ac2edf}

\bibitem[{{Rozner} {et~al.}(2023){Rozner}, {Generozov}, \& {Perets}}]{Rozner_2023}
{Rozner}, M., {Generozov}, A., \& {Perets}, H.~B. 2023, \mnras, 521, 866, \dodoi{10.1093/mnras/stad603}

\bibitem[{{Ryu} {et~al.}(2023){Ryu}, {Perna}, {Pakmor}, {Ma}, {Farmer}, \& {de Mink}}]{Ryu_2023}
{Ryu}, T., {Perna}, R., {Pakmor}, R., {et~al.} 2023, \mnras, 519, 5787, \dodoi{10.1093/mnras/stad079}

\bibitem[{{Sana} {et~al.}(2012){Sana}, {de Mink}, {de Koter}, {Langer}, {Evans}, {Gieles}, {Gosset}, {Izzard}, {Le Bouquin}, \& {Schneider}}]{Sana_2012}
{Sana}, H., {de Mink}, S.~E., {de Koter}, A., {et~al.} 2012, Science, 337, 444, \dodoi{10.1126/science.1223344}

\bibitem[{{Shu} {et~al.}(1987){Shu}, {Adams}, \& {Lizano}}]{Shu_1987}
{Shu}, F.~H., {Adams}, F.~C., \& {Lizano}, S. 1987, \araa, 25, 23, \dodoi{10.1146/annurev.aa.25.090187.000323}

\bibitem[{{Statler} {et~al.}(1987){Statler}, {Ostriker}, \& {Cohn}}]{Statler1987}
{Statler}, T.~S., {Ostriker}, J.~P., \& {Cohn}, H.~N. 1987, \apj, 316, 626, \dodoi{10.1086/165230}

\bibitem[{{Stodolkiewicz}(1986)}]{Stodolkiewicz1986}
{Stodolkiewicz}, J.~S. 1986, \actaa, 36, 19

\bibitem[{{Stoer} \& {Bulirsch}(1980)}]{stoer1980}
{Stoer}, J., \& {Bulirsch}, R. 1980, {Introduction to Numerical Analysis} ({Springer-Verlag}, New York), 430, \dodoi{https://doi.org/10.1007/978-0-387-21738-3}

\bibitem[{{Tanikawa} {et~al.}(2013){Tanikawa}, {Heggie}, {Hut}, \& {Makino}}]{Tanikawa_2013}
{Tanikawa}, A., {Heggie}, D.~C., {Hut}, P., \& {Makino}, J. 2013, Astronomy and Computing, 3, 35, \dodoi{10.1016/j.ascom.2013.11.002}

\bibitem[{{Tokovinin}(2020)}]{Tokovinin_2020}
{Tokovinin}, A. 2020, \mnras, 496, 987, \dodoi{10.1093/mnras/staa1639}

\bibitem[{{Trani} {et~al.}(2019{\natexlab{a}}){Trani}, {Fujii}, \& {Spera}}]{trani2019a}
{Trani}, A.~A., {Fujii}, M.~S., \& {Spera}, M. 2019{\natexlab{a}}, \apj, 875, 42, \dodoi{10.3847/1538-4357/ab0e70}

\bibitem[{{Trani} {et~al.}(2023){Trani}, {Quaini}, \& {Colpi}}]{Trani_2023}
{Trani}, A.~A., {Quaini}, S., \& {Colpi}, M. 2023, arXiv e-prints, arXiv:2312.13281, \dodoi{10.48550/arXiv.2312.13281}

\bibitem[{{Trani} \& {Spera}(2022)}]{tsunami2022}
{Trani}, A.~A., \& {Spera}, M. 2022, arXiv e-prints, arXiv:2206.10583.
\newblock \doarXiv{2206.10583}

\bibitem[{{Trani} \& {Spera}(2023)}]{Trani2023}
{Trani}, A.~A., \& {Spera}, M. 2023, in The Predictive Power of Computational Astrophysics as a Discover Tool, ed. D.~{Bisikalo}, D.~{Wiebe}, \& C.~{Boily}, Vol. 362, 404--409, \dodoi{10.1017/S1743921322001818}

\bibitem[{{Trani} {et~al.}(2019{\natexlab{b}}){Trani}, {Spera}, {Leigh}, \& {Fujii}}]{trani2019b}
{Trani}, A.~A., {Spera}, M., {Leigh}, N. W.~C., \& {Fujii}, M.~S. 2019{\natexlab{b}}, \apj, 885, 135, \dodoi{10.3847/1538-4357/ab480a}

\bibitem[{{Wang} {et~al.}(2016){Wang}, {Spurzem}, {Aarseth}, {Giersz}, {Askar}, {Berczik}, {Naab}, {Schadow}, \& {Kouwenhoven}}]{Wang2016}
{Wang}, L., {Spurzem}, R., {Aarseth}, S., {et~al.} 2016, \mnras, 458, 1450, \dodoi{10.1093/mnras/stw274}

\bibitem[{{Weatherford} {et~al.}(2023){Weatherford}, {K{\i}ro{\u{g}}lu}, {Fragione}, {Chatterjee}, {Kremer}, \& {Rasio}}]{Weatherford_2023a}
{Weatherford}, N.~C., {K{\i}ro{\u{g}}lu}, F., {Fragione}, G., {et~al.} 2023, \apj, 946, 104, \dodoi{10.3847/1538-4357/acbcc1}

\bibitem[{{Xu} {et~al.}(2023){Xu}, {Hwang}, {Hamilton}, \& {Lai}}]{Xu_2023}
{Xu}, S., {Hwang}, H.-C., {Hamilton}, C., \& {Lai}, D. 2023, \apjl, 949, L28, \dodoi{10.3847/2041-8213/acd6f7}

\end{thebibliography}
\bibliographystyle{aasjournal}

\bigskip

\section*{Appendix}
Table~\ref{table:q} comprises all the 3BBF probabilities from Figure~\ref{fig:sigvseta_q}.

\begin{table*}
\caption{The 3BBF probabilities from Figure~\ref{fig:sigvseta_q} for selected cases of unequal masses, distinguished by hardness and specific mass pairing. Each sub-table contains the values for different minimum hardness ($a_{\rm ij}/a_{\rm fs, ij}$), rows contain probabilities for a specific set of mass ratios $(q_1, q_2)$ with fixed $q_0=1$, and columns specify pairing combination. Additionally, the $95\%$ confidence interval is included for every probability. There are exactly $10^9$ simulations generated per mass ratio or $2.1\times10^{10}$ in total.}

\label{table:q}
\begin{minipage}{\textwidth}
\begin{tabular}{c|cccc}
\toprule
\multicolumn{5}{c}{all} \\
$(q_1, \, \, \, \, \, q_2)$ & $P_{\rm tot}$ & $P_{01}$ & $P_{02}$ & $P_{12}$ \\
\midrule
$(0.10, 0.10)$ & $(4.05 \pm 0.00) \times 10^{-3}$ & $(9.54 \pm 0.02) \times 10^{-4}$ & $(9.53 \pm 0.02) \times 10^{-4}$ & $(2.14 \pm 0.00) \times 10^{-3}$ \\
$(0.25, 0.10)$ & $(7.04 \pm 0.01) \times 10^{-3}$ & $(1.04 \pm 0.00) \times 10^{-3}$ & $(2.53 \pm 0.00) \times 10^{-3}$ & $(3.47 \pm 0.00) \times 10^{-3}$ \\
$(0.40, 0.10)$ & $(8.78 \pm 0.01) \times 10^{-3}$ & $(1.02 \pm 0.00) \times 10^{-3}$ & $(3.60 \pm 0.00) \times 10^{-3}$ & $(4.16 \pm 0.00) \times 10^{-3}$ \\
$(0.55, 0.10)$ & $(9.64 \pm 0.01) \times 10^{-3}$ & $(9.44 \pm 0.02) \times 10^{-4}$ & $(4.20 \pm 0.00) \times 10^{-3}$ & $(4.49 \pm 0.00) \times 10^{-3}$ \\
$(0.70, 0.10)$ & $(9.96 \pm 0.01) \times 10^{-3}$ & $(8.60 \pm 0.02) \times 10^{-4}$ & $(4.48 \pm 0.00) \times 10^{-3}$ & $(4.62 \pm 0.00) \times 10^{-3}$ \\
$(0.85, 0.10)$ & $(9.98 \pm 0.01) \times 10^{-3}$ & $(7.80 \pm 0.02) \times 10^{-4}$ & $(4.57 \pm 0.00) \times 10^{-3}$ & $(4.62 \pm 0.00) \times 10^{-3}$ \\
$(1.00, 0.10)$ & $(9.82 \pm 0.01) \times 10^{-3}$ & $(7.09 \pm 0.02) \times 10^{-4}$ & $(4.55 \pm 0.00) \times 10^{-3}$ & $(4.55 \pm 0.00) \times 10^{-3}$ \\
\hline
$(0.33, 0.33)$ & $(1.73 \pm 0.00) \times 10^{-2}$ & $(4.82 \pm 0.00) \times 10^{-3}$ & $(4.78 \pm 0.00) \times 10^{-3}$ & $(7.69 \pm 0.01) \times 10^{-3}$ \\
$(0.44, 0.33)$ & $(1.86 \pm 0.00) \times 10^{-2}$ & $(4.71 \pm 0.00) \times 10^{-3}$ & $(5.80 \pm 0.00) \times 10^{-3}$ & $(8.05 \pm 0.01) \times 10^{-3}$ \\
$(0.56, 0.33)$ & $(1.90 \pm 0.00) \times 10^{-2}$ & $(4.46 \pm 0.00) \times 10^{-3}$ & $(6.49 \pm 0.00) \times 10^{-3}$ & $(8.07 \pm 0.01) \times 10^{-3}$ \\
$(0.67, 0.33)$ & $(1.89 \pm 0.00) \times 10^{-2}$ & $(4.19 \pm 0.00) \times 10^{-3}$ & $(6.86 \pm 0.01) \times 10^{-3}$ & $(7.90 \pm 0.01) \times 10^{-3}$ \\
$(0.78, 0.33)$ & $(1.86 \pm 0.00) \times 10^{-2}$ & $(3.90 \pm 0.00) \times 10^{-3}$ & $(7.03 \pm 0.01) \times 10^{-3}$ & $(7.64 \pm 0.01) \times 10^{-3}$ \\
$(0.89, 0.33)$ & $(1.81 \pm 0.00) \times 10^{-2}$ & $(3.64 \pm 0.00) \times 10^{-3}$ & $(7.08 \pm 0.01) \times 10^{-3}$ & $(7.36 \pm 0.01) \times 10^{-3}$ \\
$(1.00, 0.33)$ & $(1.75 \pm 0.00) \times 10^{-2}$ & $(3.39 \pm 0.00) \times 10^{-3}$ & $(7.04 \pm 0.01) \times 10^{-3}$ & $(7.04 \pm 0.01) \times 10^{-3}$ \\
\hline
$(0.50, 0.50)$ & $(2.49 \pm 0.00) \times 10^{-2}$ & $(7.38 \pm 0.01) \times 10^{-3}$ & $(7.38 \pm 0.01) \times 10^{-3}$ & $(1.02 \pm 0.00) \times 10^{-2}$ \\
$(0.58, 0.50)$ & $(2.51 \pm 0.00) \times 10^{-2}$ & $(7.14 \pm 0.01) \times 10^{-3}$ & $(7.90 \pm 0.01) \times 10^{-3}$ & $(1.01 \pm 0.00) \times 10^{-2}$ \\
$(0.67, 0.50)$ & $(2.50 \pm 0.00) \times 10^{-2}$ & $(6.84 \pm 0.01) \times 10^{-3}$ & $(8.29 \pm 0.01) \times 10^{-3}$ & $(9.87 \pm 0.01) \times 10^{-3}$ \\
$(0.75, 0.50)$ & $(2.47 \pm 0.00) \times 10^{-2}$ & $(6.55 \pm 0.01) \times 10^{-3}$ & $(8.50 \pm 0.01) \times 10^{-3}$ & $(9.62 \pm 0.01) \times 10^{-3}$ \\
$(0.83, 0.50)$ & $(2.42 \pm 0.00) \times 10^{-2}$ & $(6.25 \pm 0.00) \times 10^{-3}$ & $(8.63 \pm 0.01) \times 10^{-3}$ & $(9.33 \pm 0.01) \times 10^{-3}$ \\
$(0.92, 0.50)$ & $(2.36 \pm 0.00) \times 10^{-2}$ & $(5.92 \pm 0.00) \times 10^{-3}$ & $(8.68 \pm 0.01) \times 10^{-3}$ & $(8.98 \pm 0.01) \times 10^{-3}$ \\
$(1.00, 0.50)$ & $(2.30 \pm 0.00) \times 10^{-2}$ & $(5.65 \pm 0.00) \times 10^{-3}$ & $(8.67 \pm 0.01) \times 10^{-3}$ & $(8.67 \pm 0.01) \times 10^{-3}$ \\
\bottomrule
\end{tabular}

\end{minipage}

\begin{minipage}{\textwidth}

\begin{tabular}{c|cccc}
\toprule
\multicolumn{5}{c}{$a_{\rm ij}/a_{\rm fs, ij}<10$}\\
$(q_1, \, \, \, \, \, q_2)$ & $P_{\rm tot}$ & $P_{01}$ & $P_{02}$ & $P_{12}$ \\
\midrule
$(0.10, 0.10)$ & $(3.03 \pm 0.03) \times 10^{-5}$ & $(1.41 \pm 0.02) \times 10^{-5}$ & $(1.42 \pm 0.02) \times 10^{-5}$ & $(1.97 \pm 0.09) \times 10^{-6}$ \\
$(0.25, 0.10)$ & $(8.70 \pm 0.06) \times 10^{-5}$ & $(1.06 \pm 0.02) \times 10^{-5}$ & $(6.78 \pm 0.05) \times 10^{-5}$ & $(8.55 \pm 0.18) \times 10^{-6}$ \\
$(0.40, 0.10)$ & $(1.34 \pm 0.01) \times 10^{-4}$ & $(7.79 \pm 0.17) \times 10^{-6}$ & $(1.07 \pm 0.01) \times 10^{-4}$ & $(1.89 \pm 0.03) \times 10^{-5}$ \\
$(0.55, 0.10)$ & $(1.50 \pm 0.01) \times 10^{-4}$ & $(5.53 \pm 0.15) \times 10^{-6}$ & $(1.13 \pm 0.01) \times 10^{-4}$ & $(3.12 \pm 0.03) \times 10^{-5}$ \\
$(0.70, 0.10)$ & $(1.49 \pm 0.01) \times 10^{-4}$ & $(3.98 \pm 0.12) \times 10^{-6}$ & $(1.01 \pm 0.01) \times 10^{-4}$ & $(4.44 \pm 0.04) \times 10^{-5}$ \\
$(0.85, 0.10)$ & $(1.43 \pm 0.01) \times 10^{-4}$ & $(3.02 \pm 0.11) \times 10^{-6}$ & $(8.39 \pm 0.06) \times 10^{-5}$ & $(5.63 \pm 0.05) \times 10^{-5}$ \\
$(1.00, 0.10)$ & $(1.39 \pm 0.01) \times 10^{-4}$ & $(2.34 \pm 0.10) \times 10^{-6}$ & $(6.80 \pm 0.05) \times 10^{-5}$ & $(6.84 \pm 0.05) \times 10^{-5}$ \\
\hline
$(0.33, 0.33)$ & $(3.69 \pm 0.01) \times 10^{-4}$ & $(1.52 \pm 0.01) \times 10^{-4}$ & $(1.50 \pm 0.01) \times 10^{-4}$ & $(6.75 \pm 0.05) \times 10^{-5}$ \\
$(0.44, 0.33)$ & $(4.08 \pm 0.01) \times 10^{-4}$ & $(1.35 \pm 0.01) \times 10^{-4}$ & $(1.89 \pm 0.01) \times 10^{-4}$ & $(8.44 \pm 0.06) \times 10^{-5}$ \\
$(0.56, 0.33)$ & $(4.17 \pm 0.01) \times 10^{-4}$ & $(1.18 \pm 0.01) \times 10^{-4}$ & $(2.00 \pm 0.01) \times 10^{-4}$ & $(9.96 \pm 0.06) \times 10^{-5}$ \\
$(0.67, 0.33)$ & $(4.07 \pm 0.01) \times 10^{-4}$ & $(1.04 \pm 0.01) \times 10^{-4}$ & $(1.92 \pm 0.01) \times 10^{-4}$ & $(1.12 \pm 0.01) \times 10^{-4}$ \\
$(0.78, 0.33)$ & $(3.90 \pm 0.01) \times 10^{-4}$ & $(9.11 \pm 0.06) \times 10^{-5}$ & $(1.76 \pm 0.01) \times 10^{-4}$ & $(1.23 \pm 0.01) \times 10^{-4}$ \\
$(0.89, 0.33)$ & $(3.67 \pm 0.01) \times 10^{-4}$ & $(8.05 \pm 0.06) \times 10^{-5}$ & $(1.56 \pm 0.01) \times 10^{-4}$ & $(1.31 \pm 0.01) \times 10^{-4}$ \\
$(1.00, 0.33)$ & $(3.46 \pm 0.01) \times 10^{-4}$ & $(7.14 \pm 0.05) \times 10^{-5}$ & $(1.37 \pm 0.01) \times 10^{-4}$ & $(1.37 \pm 0.01) \times 10^{-4}$ \\
\hline
$(0.50, 0.50)$ & $(7.05 \pm 0.02) \times 10^{-4}$ & $(2.65 \pm 0.01) \times 10^{-4}$ & $(2.65 \pm 0.01) \times 10^{-4}$ & $(1.76 \pm 0.01) \times 10^{-4}$ \\
$(0.58, 0.50)$ & $(7.02 \pm 0.02) \times 10^{-4}$ & $(2.46 \pm 0.01) \times 10^{-4}$ & $(2.72 \pm 0.01) \times 10^{-4}$ & $(1.85 \pm 0.01) \times 10^{-4}$ \\
$(0.67, 0.50)$ & $(6.84 \pm 0.02) \times 10^{-4}$ & $(2.24 \pm 0.01) \times 10^{-4}$ & $(2.68 \pm 0.01) \times 10^{-4}$ & $(1.92 \pm 0.01) \times 10^{-4}$ \\
$(0.75, 0.50)$ & $(6.62 \pm 0.02) \times 10^{-4}$ & $(2.09 \pm 0.01) \times 10^{-4}$ & $(2.56 \pm 0.01) \times 10^{-4}$ & $(1.96 \pm 0.01) \times 10^{-4}$ \\
$(0.83, 0.50)$ & $(6.34 \pm 0.02) \times 10^{-4}$ & $(1.93 \pm 0.01) \times 10^{-4}$ & $(2.42 \pm 0.01) \times 10^{-4}$ & $(2.00 \pm 0.01) \times 10^{-4}$ \\
$(0.92, 0.50)$ & $(6.00 \pm 0.02) \times 10^{-4}$ & $(1.77 \pm 0.01) \times 10^{-4}$ & $(2.21 \pm 0.01) \times 10^{-4}$ & $(2.02 \pm 0.01) \times 10^{-4}$ \\
$(1.00, 0.50)$ & $(5.69 \pm 0.01) \times 10^{-4}$ & $(1.64 \pm 0.01) \times 10^{-4}$ & $(2.03 \pm 0.01) \times 10^{-4}$ & $(2.03 \pm 0.01) \times 10^{-4}$ \\
\bottomrule
\end{tabular}

\end{minipage}
\end{table*}

\begin{table*}
\centering
\text{\textbf{Table~\ref{table:q}} (Continued)}

\begin{minipage}{\textwidth}
\begin{tabular}{c|cccc}
\toprule
\multicolumn{5}{c}{$a_{\rm ij}/a_{\rm fs, ij}<3$} \\
$(q_1, \, \, \, \, \, q_2)$ & $P_{\rm tot}$ & $P_{01}$ & $P_{02}$ & $P_{12}$ \\
\midrule
$(0.10, 0.10)$ & $(1.66 \pm 0.08) \times 10^{-6}$ & $(8.21 \pm 0.56) \times 10^{-7}$ & $(8.01 \pm 0.56) \times 10^{-7}$ & $(4.03 \pm 1.25) \times 10^{-8}$ \\
$(0.25, 0.10)$ & $(3.97 \pm 0.12) \times 10^{-6}$ & $(6.12 \pm 0.49) \times 10^{-7}$ & $(3.18 \pm 0.11) \times 10^{-6}$ & $(1.83 \pm 0.27) \times 10^{-7}$ \\
$(0.40, 0.10)$ & $(5.34 \pm 0.14) \times 10^{-6}$ & $(4.32 \pm 0.41) \times 10^{-7}$ & $(4.44 \pm 0.13) \times 10^{-6}$ & $(4.66 \pm 0.42) \times 10^{-7}$ \\
$(0.55, 0.10)$ & $(5.24 \pm 0.14) \times 10^{-6}$ & $(2.75 \pm 0.33) \times 10^{-7}$ & $(4.18 \pm 0.13) \times 10^{-6}$ & $(7.83 \pm 0.55) \times 10^{-7}$ \\
$(0.70, 0.10)$ & $(4.62 \pm 0.13) \times 10^{-6}$ & $(1.98 \pm 0.28) \times 10^{-7}$ & $(3.27 \pm 0.11) \times 10^{-6}$ & $(1.15 \pm 0.07) \times 10^{-6}$ \\
$(0.85, 0.10)$ & $(4.26 \pm 0.13) \times 10^{-6}$ & $(1.68 \pm 0.26) \times 10^{-7}$ & $(2.50 \pm 0.10) \times 10^{-6}$ & $(1.59 \pm 0.08) \times 10^{-6}$ \\
$(1.00, 0.10)$ & $(3.92 \pm 0.12) \times 10^{-6}$ & $(1.25 \pm 0.22) \times 10^{-7}$ & $(1.84 \pm 0.08) \times 10^{-6}$ & $(1.95 \pm 0.09) \times 10^{-6}$ \\
\hline
$(0.33, 0.33)$ & $(1.69 \pm 0.03) \times 10^{-5}$ & $(7.52 \pm 0.17) \times 10^{-6}$ & $(7.50 \pm 0.17) \times 10^{-6}$ & $(1.84 \pm 0.08) \times 10^{-6}$ \\
$(0.44, 0.33)$ & $(1.75 \pm 0.03) \times 10^{-5}$ & $(6.67 \pm 0.16) \times 10^{-6}$ & $(8.48 \pm 0.18) \times 10^{-6}$ & $(2.36 \pm 0.10) \times 10^{-6}$ \\
$(0.56, 0.33)$ & $(1.66 \pm 0.03) \times 10^{-5}$ & $(5.80 \pm 0.15) \times 10^{-6}$ & $(8.06 \pm 0.18) \times 10^{-6}$ & $(2.79 \pm 0.10) \times 10^{-6}$ \\
$(0.67, 0.33)$ & $(1.55 \pm 0.02) \times 10^{-5}$ & $(5.20 \pm 0.14) \times 10^{-6}$ & $(7.17 \pm 0.17) \times 10^{-6}$ & $(3.17 \pm 0.11) \times 10^{-6}$ \\
$(0.78, 0.33)$ & $(1.44 \pm 0.02) \times 10^{-5}$ & $(4.70 \pm 0.13) \times 10^{-6}$ & $(6.14 \pm 0.15) \times 10^{-6}$ & $(3.58 \pm 0.12) \times 10^{-6}$ \\
$(0.89, 0.33)$ & $(1.31 \pm 0.02) \times 10^{-5}$ & $(4.10 \pm 0.13) \times 10^{-6}$ & $(5.12 \pm 0.14) \times 10^{-6}$ & $(3.91 \pm 0.12) \times 10^{-6}$ \\
$(1.00, 0.33)$ & $(1.21 \pm 0.02) \times 10^{-5}$ & $(3.69 \pm 0.12) \times 10^{-6}$ & $(4.15 \pm 0.13) \times 10^{-6}$ & $(4.21 \pm 0.13) \times 10^{-6}$ \\
\hline
$(0.50, 0.50)$ & $(2.94 \pm 0.03) \times 10^{-5}$ & $(1.21 \pm 0.02) \times 10^{-5}$ & $(1.23 \pm 0.02) \times 10^{-5}$ & $(5.07 \pm 0.14) \times 10^{-6}$ \\
$(0.58, 0.50)$ & $(2.90 \pm 0.03) \times 10^{-5}$ & $(1.16 \pm 0.02) \times 10^{-5}$ & $(1.19 \pm 0.02) \times 10^{-5}$ & $(5.50 \pm 0.15) \times 10^{-6}$ \\
$(0.67, 0.50)$ & $(2.71 \pm 0.03) \times 10^{-5}$ & $(1.04 \pm 0.02) \times 10^{-5}$ & $(1.09 \pm 0.02) \times 10^{-5}$ & $(5.81 \pm 0.15) \times 10^{-6}$ \\
$(0.75, 0.50)$ & $(2.57 \pm 0.03) \times 10^{-5}$ & $(9.89 \pm 0.20) \times 10^{-6}$ & $(9.81 \pm 0.19) \times 10^{-6}$ & $(6.02 \pm 0.15) \times 10^{-6}$ \\
$(0.83, 0.50)$ & $(2.42 \pm 0.03) \times 10^{-5}$ & $(9.23 \pm 0.19) \times 10^{-6}$ & $(8.76 \pm 0.18) \times 10^{-6}$ & $(6.23 \pm 0.16) \times 10^{-6}$ \\
$(0.92, 0.50)$ & $(2.27 \pm 0.03) \times 10^{-5}$ & $(8.53 \pm 0.18) \times 10^{-6}$ & $(7.63 \pm 0.17) \times 10^{-6}$ & $(6.50 \pm 0.16) \times 10^{-6}$ \\
$(1.00, 0.50)$ & $(2.13 \pm 0.03) \times 10^{-5}$ & $(8.01 \pm 0.18) \times 10^{-6}$ & $(6.57 \pm 0.16) \times 10^{-6}$ & $(6.69 \pm 0.16) \times 10^{-6}$ \\
\bottomrule
\end{tabular}

\end{minipage}

\begin{minipage}{\textwidth}

\begin{tabular}{c|cccc}
\toprule
\multicolumn{5}{c}{$a_{\rm ij}/a_{\rm fs, ij}<1$} \\
$(q_1, \, \, \, \, \, q_2)$ & $P_{\rm tot}$ & $P_{01}$ & $P_{02}$ & $P_{12}$ \\
\midrule
$(0.10, 0.10)$ & $(4.63 \pm 1.34) \times 10^{-8}$ & $(2.12 \pm 0.90) \times 10^{-8}$ & $(2.32 \pm 0.95) \times 10^{-8}$ & $(2.02 \pm 2.79) \times 10^{-9}$ \\
$(0.25, 0.10)$ & $(1.60 \pm 0.25) \times 10^{-7}$ & $(3.02 \pm 1.08) \times 10^{-8}$ & $(1.26 \pm 0.22) \times 10^{-7}$ & $(4.03 \pm 3.95) \times 10^{-9}$ \\
$(0.40, 0.10)$ & $(1.81 \pm 0.26) \times 10^{-7}$ & $(1.71 \pm 0.81) \times 10^{-8}$ & $(1.49 \pm 0.24) \times 10^{-7}$ & $(1.51 \pm 0.76) \times 10^{-8}$ \\
$(0.55, 0.10)$ & $(1.86 \pm 0.27) \times 10^{-7}$ & $(2.01 \pm 0.88) \times 10^{-8}$ & $(1.41 \pm 0.23) \times 10^{-7}$ & $(2.52 \pm 0.99) \times 10^{-8}$ \\
$(0.70, 0.10)$ & $(1.39 \pm 0.23) \times 10^{-7}$ & $(1.21 \pm 0.68) \times 10^{-8}$ & $(1.01 \pm 0.20) \times 10^{-7}$ & $(2.62 \pm 1.01) \times 10^{-8}$ \\
$(0.85, 0.10)$ & $(1.17 \pm 0.21) \times 10^{-7}$ & $(4.03 \pm 3.95) \times 10^{-9}$ & $(7.35 \pm 1.69) \times 10^{-8}$ & $(3.93 \pm 1.23) \times 10^{-8}$ \\
$(1.00, 0.10)$ & $(1.20 \pm 0.22) \times 10^{-7}$ & $(9.07 \pm 5.92) \times 10^{-9}$ & $(5.74 \pm 1.49) \times 10^{-8}$ & $(5.34 \pm 1.44) \times 10^{-8}$ \\
\hline
$(0.33, 0.33)$ & $(5.94 \pm 0.48) \times 10^{-7}$ & $(2.56 \pm 0.32) \times 10^{-7}$ & $(2.86 \pm 0.33) \times 10^{-7}$ & $(5.14 \pm 1.41) \times 10^{-8}$ \\
$(0.44, 0.33)$ & $(5.99 \pm 0.48) \times 10^{-7}$ & $(2.47 \pm 0.31) \times 10^{-7}$ & $(2.97 \pm 0.34) \times 10^{-7}$ & $(5.55 \pm 1.47) \times 10^{-8}$ \\
$(0.56, 0.33)$ & $(5.48 \pm 0.46) \times 10^{-7}$ & $(2.11 \pm 0.29) \times 10^{-7}$ & $(2.56 \pm 0.31) \times 10^{-7}$ & $(8.07 \pm 1.77) \times 10^{-8}$ \\
$(0.67, 0.33)$ & $(5.15 \pm 0.45) \times 10^{-7}$ & $(1.95 \pm 0.27) \times 10^{-7}$ & $(2.38 \pm 0.30) \times 10^{-7}$ & $(8.27 \pm 1.79) \times 10^{-8}$ \\
$(0.78, 0.33)$ & $(4.92 \pm 0.44) \times 10^{-7}$ & $(1.87 \pm 0.27) \times 10^{-7}$ & $(1.82 \pm 0.27) \times 10^{-7}$ & $(1.23 \pm 0.22) \times 10^{-7}$ \\
$(0.89, 0.33)$ & $(4.35 \pm 0.41) \times 10^{-7}$ & $(1.62 \pm 0.25) \times 10^{-7}$ & $(1.35 \pm 0.23) \times 10^{-7}$ & $(1.37 \pm 0.23) \times 10^{-7}$ \\
$(1.00, 0.33)$ & $(3.88 \pm 0.39) \times 10^{-7}$ & $(1.31 \pm 0.23) \times 10^{-7}$ & $(1.41 \pm 0.23) \times 10^{-7}$ & $(1.16 \pm 0.21) \times 10^{-7}$ \\
\hline
$(0.50, 0.50)$ & $(9.70 \pm 0.61) \times 10^{-7}$ & $(4.23 \pm 0.40) \times 10^{-7}$ & $(4.11 \pm 0.40) \times 10^{-7}$ & $(1.36 \pm 0.23) \times 10^{-7}$ \\
$(0.58, 0.50)$ & $(9.49 \pm 0.61) \times 10^{-7}$ & $(3.87 \pm 0.39) \times 10^{-7}$ & $(4.04 \pm 0.40) \times 10^{-7}$ & $(1.57 \pm 0.25) \times 10^{-7}$ \\
$(0.67, 0.50)$ & $(8.64 \pm 0.58) \times 10^{-7}$ & $(3.93 \pm 0.39) \times 10^{-7}$ & $(3.28 \pm 0.36) \times 10^{-7}$ & $(1.43 \pm 0.24) \times 10^{-7}$ \\
$(0.75, 0.50)$ & $(8.21 \pm 0.56) \times 10^{-7}$ & $(3.77 \pm 0.38) \times 10^{-7}$ & $(2.74 \pm 0.33) \times 10^{-7}$ & $(1.69 \pm 0.26) \times 10^{-7}$ \\
$(0.83, 0.50)$ & $(8.18 \pm 0.56) \times 10^{-7}$ & $(3.44 \pm 0.37) \times 10^{-7}$ & $(2.79 \pm 0.33) \times 10^{-7}$ & $(1.95 \pm 0.27) \times 10^{-7}$ \\
$(0.92, 0.50)$ & $(7.60 \pm 0.54) \times 10^{-7}$ & $(3.42 \pm 0.36) \times 10^{-7}$ & $(2.44 \pm 0.31) \times 10^{-7}$ & $(1.74 \pm 0.26) \times 10^{-7}$ \\
$(1.00, 0.50)$ & $(6.87 \pm 0.52) \times 10^{-7}$ & $(2.85 \pm 0.33) \times 10^{-7}$ & $(1.92 \pm 0.27) \times 10^{-7}$ & $(2.10 \pm 0.29) \times 10^{-7}$ \\
\bottomrule
\end{tabular}

\end{minipage}
\end{table*}



\end{document}